\documentclass[11pt,onecolumn]{article}
\usepackage[utf8]{inputenc} 
\usepackage[T1]{fontenc}    
\usepackage{amsfonts}       
\usepackage{times}
\usepackage{amsmath}
\usepackage{mathrsfs}
\usepackage{amssymb}
\usepackage{theorem}
\usepackage{graphicx}
\usepackage{subfigure}
\usepackage{multicol}
\usepackage{color}
\usepackage{xspace} 
\usepackage{algorithm}
\usepackage{algorithmic}
\usepackage{soul}
\usepackage{authblk}
\usepackage{hyperref}

\definecolor{darkred}{rgb}{1, 0.1, 0.3}
\definecolor{darkgreen}{rgb}{0.5, 0.8, 0.1}
\definecolor{darkpurple}{rgb}{1.0, 0, 1.0}
\definecolor{darkblue}{rgb}{0, 0, 1.0}

\newtheorem{definition}{Definition}

\newcommand{\HGCN}      {{\sf HS-GNN}}

\newcommand{\atCNT}     {{\texttt S}}

\newcommand{\PI}        {{\mathrm{PI}}}

\newcommand{\dgm}           {\mathrm{Dg}}
\newcommand{\reals}         {\mathbb{R}}
\newcommand{\aD}            {{\mathrm{D}}}

\title{Prediction of Carbon Nanostructure Mechanical Properties and Role of Defects Using Machine Learning}
\author[1]{Qi Zhao}
\author[2]{Jordan J. Winetrout}
\author[4]{Yanxun Xu}
\author[1,3]{Yusu Wang\thanks{yusuwang@ucsd.edu}}
\author[2]{Hendrik Heinz\thanks{hendrik.heinz@colorado.edu}}
\affil[1]{Department of Computer Science and Engineering, University of California San Diego, La Jolla, CA 92093}
\affil[3]{Halıcıoğlu Data Science Institute, University of California San Diego, La Jolla, CA 92093}
\affil[2]{Department of Chemical and Biological Engineering, University of Colorado Boulder, Boulder, CO 80309, and Materials Science and Engineering Program, University of Colorado Boulder, Boulder, CO 80309}
\affil[4]{Department of Applied Mathematics and Statistics, Johns Hopkins University, Baltimore, MD 21218}

\date{September 2021}
\begin{document}

\maketitle
\begin{abstract}
Carbon fiber and graphene-based nanostructures such as carbon nanotubes (CNTs) and defective structures have extraordinary potential as strong and lightweight materials. A longstanding bottleneck has been lack of understanding and implementation of atomic-scale engineering to harness the theoretical limits of modulus and tensile strength, of which only a fraction is routinely reached today. Here we demonstrate accurate and fast predictions of mechanical properties for CNTs and arbitrary 3D graphitic assemblies based on a training set of over 1000 stress-strain curves from cutting-edge reactive MD simulation and machine learning (ML). Several ML methods are compared and show that our newly proposed hierarchically structured graph neural networks with spatial information (\HGCN{}s) achieve predictions in modulus and strength for any 3D nanostructure with only 5-10\% error across a wide range of possible values. The reliability is sufficient for practical applications and a great improvement over off-the shelf ML methods with up to 50\% deviation, as well as over earlier models for specific chemistry with ~20\% deviation. The algorithms allow more than 10 times faster mechanical property predictions than traditional molecular dynamics simulations, the identification of the role of defects and random 3D morphology, and high-throughput screening of 3D structures for enhanced mechanical properties. The algorithms can potentially be scaled to morphologies up to 100 nm in size, expanded for chemically similar compounds, and trained to predict a broader range of properties. 
\end{abstract}

\maketitle

\section{Introduction}
Carbon fiber is among the most promising engineering materials for the 21st century due to superior tensile strength and modulus relative to low weight (Figure 1).\cite{kumar2017review} It is extensively used in the automotive, aviation, and aerospace industry (Figure 1a). Carbon fiber is composed of nanostructured carbon such as carbon nanotubes (CNTs), graphitic layers, and sometimes polymer binders, for example, in high-strength carbon fiber yarns. In addition to structural applications and fabricating ultra-strong composites, CNTs and graphitic materials are also used in catalyst supports, ultra-small electrical and thermal conductors, as well as membranes for water desalination that benefit from controlled microstructure, conductivity, and mechanical stability.\cite{Merlet2012_C_supercap}\cite{Koenig2012_graph_sieve}

One of the major challenges is closing the gap between the current mechanical properties, which are already superior to most other materials, and the theoretically possible performance, which is significantly higher (Figure 1b).\cite{baiCNT2018_80GPa} Specifically, the theoretically possible Young's Modulus is ~1 TPa while current IM-7 fiber achieves 400 GPa, and the theoretically possible tensile strength amounts to ~100 GPa while current IM-7 fiber barely reaches 8 GPa.\cite{kumar2017review} There is a space of improvements by multiples to explore, which can facilitate revolutionary changes in multiple industries including automotive, air and space flight.

Since the discovery of graphene and CNTs, there has been a multidisciplinary effort to understand and utilize their unique and remarkable properties. Loss in mechanical performance, in particular, is related to defects, impurities, random orientations as well as finite lengths and resulting gaps between CNTs.\cite{baiCNT2018_80GPa} It remains difficult to quantify the impact of these features on mechanical properties and tackle related synthesis and processing challenges towards customized manufacturing from the atomic scale to the microscopic scale. Characterization of the structure in experiments relies on X-ray scattering, microscopy and 3D tomographic reconstruction, as well as indirect methods such as Raman spectroscopy.\cite{jolowskyCNT2018_Liang} Current measurements remain tedious and expensive, limited in the number of samples and in revealing failure mechanisms down to the atomic and nanometer-scale. To complement such efforts, simulation methods have been employed such as molecular dynamics simulations and density functional theory. The advantages of these techniques include insight at the scale of atoms, access to the large nanometer scale, and the ability to simulate entire stress-strain experiments from equilibrium to failure in high accuracy.\cite{pramanik2019_CNT_PAN}\cite{winetrout2021iffr} Simulations then  allow to design and screen a larger number of model structures with a variety of defects and nanoscale features of interest, inspired by experimental data and theory, and examine the failure mechanisms in depth. However, although more efficient than experiments, the computational cost is still considerable.

Recently, there has been a major trend to explore properties of materials using rapidly developing techniques from computer science and machine learning (Figure 1c, d).\cite{schleder2019_ML_for_Mat} Machine learning has already achieved remarkable success on a wide range of applications, such as image classification and segmentation, language translation, and DNA sequence analysis \cite{krizhevsky2012imagenet,he2016deep,graves2013speech,vaswani2017attention,alipanahi2015predicting,nguyen2016dna,jordan2015machine}.
Different from conventional approaches, machine learning (ML) algorithms build models and infer knowledge about materials from sampled data called training data. The trained models can then be used to predict materials properties or changes in the dynamics of fictive materials structures. 
Researchers first focused on using vectorized global representations of molecular structures to predict their properties. Besides some simple features describing basic properties of molecular structures like the number of atoms or bonds, a list of molecule representation methods including Hamiltonians \cite{todeschini2008handbook}, Coulomb matrix \cite{rupp2012fast}, Bag of Bonds (BoB) \cite{hansen2015machine}, electronic density \cite{hirn2017wavelet}, symmetry functions \cite{bartok2013representing}, and fingerprints \cite{duvenaud2015convolutional} have been suggested. Kernel methods and Gaussian Models are commonly applied with those global representations in downstream machine learning tasks \cite{seko2017representation,rupp2012fast,hansen2015machine}. Kernel ridge regression and Gaussian process regression on such representations can be used to predict cohesive energies of molecular structures and other properties \cite{de2016comparing}. Although kernel methods are powerful when dealing with relatively small datasets of vectorized representations, it still has two main drawbacks. First, when the size of training data set grows increasingly large, the computation and storage kernel methods become extremely expensive. For example, the computation of a millions dimensional kernel matrix is a disaster. Second, as these vectorized representations cannot extract all major information from molecules, hidden features that are potentially important will be lost.

A more powerful machine learning framework in the age of data is deep learning (DL) \cite{lecun2015deep,goodfellow2016deep}. Roughly speaking, deep learning uses multiple neural network layers to progressively learn representations, and is often used to model the map from input raw data to output target properties. Neural networks can learn molecules representations and apply them to prediction tasks in a more intelligent and data-driven way. Depending on how they represent molecular structures, we can classify deep learning approaches as follows. 
First, SMILES \cite{weininger1989smiles} is a traditional text-processing architecture that represents molecules as sequences, on which it has been common for researchers to apply Recurrent Neural Networks (RNN) or Long Short-term Memory (LSTM) techniques \cite{jastrzkebski2016learning,xu2017seq2seq,sanchez2018inverse}.
Second, to better encode the 3D structural information of molecules, one could model a molecular structure as a 3D point cloud where each atom is represented by a point in $\mathbb{R}^3$, and then apply the PointNet \cite{qi2017pointnet} or RS-CNN \cite{liu2019relation} architecture to perform property prediction over such point cloud representation \cite{defever2019generalized}. However, simply representing molecules as point clouds can lose significant structural information such as chemical bonds. \cite{xu2019a2} and \cite{gebauer2019symmetry} then improved this approach by designing more expressive point cloud convolution networks which align local neighbor information and symmetry factorization of point distributions. 
Furthermore, fingerprints \cite{duvenaud2015convolutional,kearnes2016molecular,coley2017convolutional} have the power to encode the chemical bond relations which can be viewed as the start of applications of Graph Neural Networks for the prediction of properties of molecular structures.

Finally, a more natural way to organize molecules is to represent them as graphs (Figure 1c). Atoms are represented as nodes in graphs and two atoms are connected when there exists a chemical bond between them. Graph Neural Networks (GNNs) like Graph Convolution Networks (GCN) \cite{kipf2016semi} and Graph Attention Networks (GAT) \cite{velivckovic2017graph} can perform convolution over graph-structured data.
MPNN \cite{gilmer2017neural} has built a unified framework on designing GNNs as well as a specific architecture for predicting molecular properties. DimeNet \cite{klicpera2020directional} and D-MPNN \cite{yang2019analyzing} further developed this approach by incorporating bonded interactions into models. Although GNNs outperform other approaches mentioned above in tasks 
like classification of small molecules, there are limitations in existing approaches: (1) In addition to atom connectivity and bonded interactions, a good model should also take non-bonded relations and features characterizing spatial (geometric) patterns of atomic structures into consideration. (2) Due to over-smoothing or over-squashing phenomena, long-range interactions / signals are hard to capture, making it challenging to process large scale molecular structures and assemblies. 

\paragraph{Contribution of this paper.} 
In this paper, we develop a new machine learning framework to predict mechanical properties for carbon nanostructures directly from their atomic structure configurations (Figure 1d). Our ML framework, called \HGCN{}, is an enhanced graph neural network architecture, carefully designed to process CNT atomic structures more effectively. In particular, our \HGCN{} provides a much more powerful way to encode the input microstructure configuration than previous approaches: (i) The input is modeled as a heterogenuous graph so as to capture both the covalent bonds and the short-range (non-bonded) interactions among atoms. (ii) The geometric shape of the input structure (e.g, how the CNT sheet is curved, and what geometric patterns it forms with atoms from neighboring sheets) can affect the mechanical properties of the structure. We leverage local geometric and topological features to better encode the spatial geometric shape of the input structure. (iii) We deploy a hierarchical neural network design to capture large-scale interaction among different parts of CNT composite. 

We train our \HGCN{} on initial configurations of about 1000 independent CNTs. Tensile properties used as ground truth in training set are generated from their stress-strain curves of complex carbon nanostructures using molecular dynamics simulations with the thoroughly validated IFF force field that includes bond breaking (IFF-R) \cite{heinz2013IFF,pramanik2017acsnano,winetrout2021iffr}. 
The resulting \HGCN{} can predict Young's modulus and strength for new atomic arrangements (not used in training set) with high accuracy (5-10\% MSE error), including larger graphitic structures. The time it takes is less than 10\% of the computational cost of traditional molecular simulations. We also compare our \HGCN{} with six other ML approaches including the state-of-the art DimeNet \cite{klicpera2020directional}, and our new approach achieves significantly better accuracy than these other ML approaches. The ML source code and examples, as well as key data and run scripts for the training set, will be shared publicly.

\begin{figure}[htbp]
    \centering
    \includegraphics[scale=0.24]{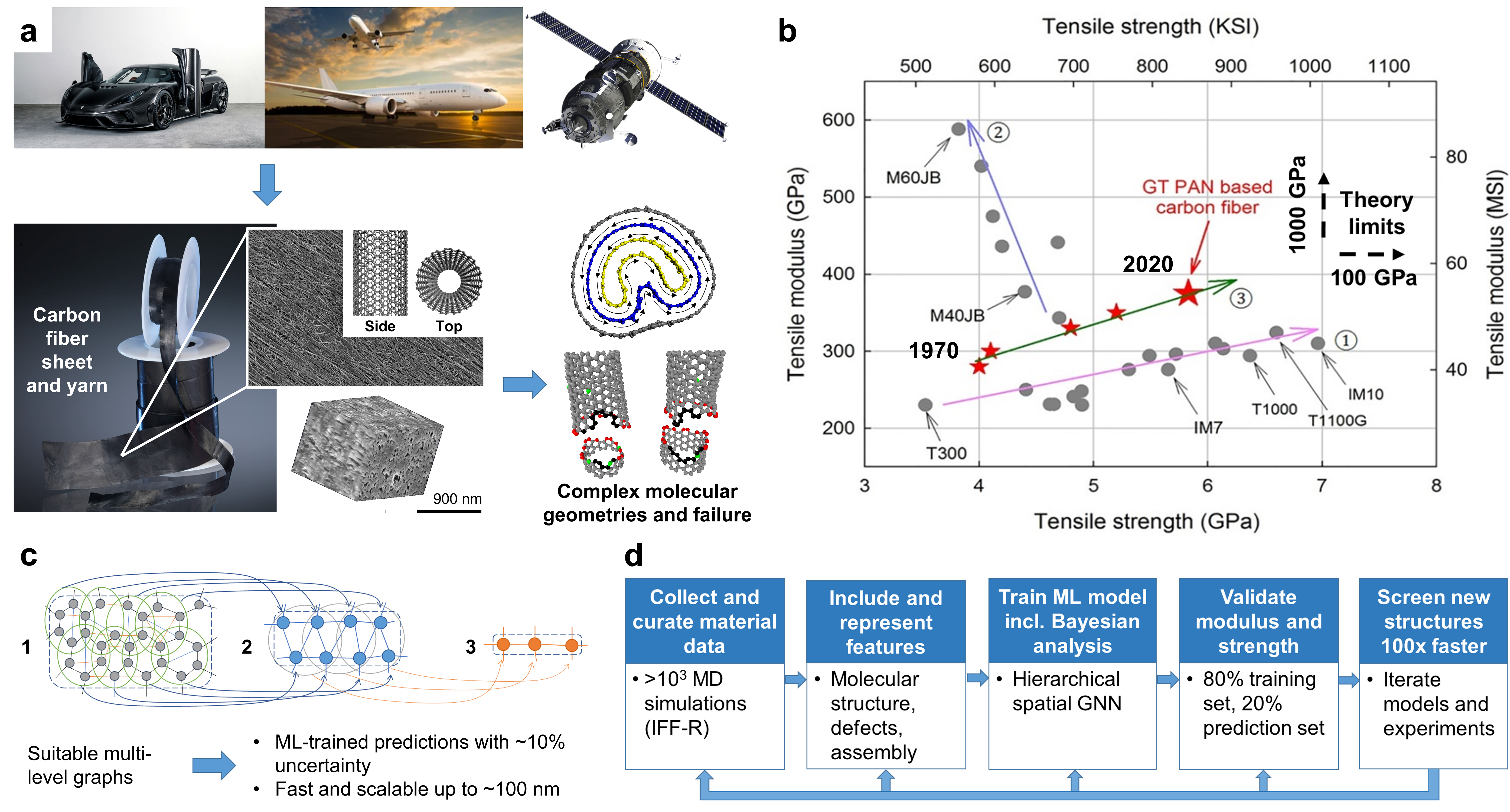}
    \caption{Hierarchical structure of carbon-based materials, challenges and new machine learning workflow to  predict mechanical properties at the nanoscale. (a) Automotive, air and space vehicles (top) increasingly require strong and lightweight materials, for which carbon fiber and related composites are a promising choice. At the microscale, carbon fiber sheet and yarn consist of aligned yet highly defective carbon fiber bundles, which can be composed of carbon nanotubes and irregular graphitic structures (bottom left). 3D imaging by tomography is typically limited to the coarse nanometer scale. In this work, we examine stress-strain relationships up to failure for a library of over 1000 structures at the small nanometer scale and feed the data into a ML model (bottom right). (b) The current performance of carbon fiber is much lower than the theoretical limits, which are ca. 1000 GPa in tensile modulus and ca. 100 GPa in tensile strength. While efforts over the last 50 years have brought steady improvements, the bottleneck is understanding and engineering of defects in the materials that originate from the molecular scale to the nanoscale. (c) We introduce a hierarchical spatial graph neural network (\HGCN{}) that shows excellent performance in mechanical property predictions for a wide range of graphitic morphologies up to the large nanometer scale. (d) The workflow includes data collection by reactive MD simulations (validated relative to experiment), representation of the structures by the \HGCN{}, training of the ML model with the data for  modulus and strength, and fast screening (prediction) of mechanical properties of unknown structures to quantify the role of defects and desirable new nanoscale designs. 
     }
    \label{Figure1}
\end{figure}
\clearpage
\section{Results and Discussion}

\subsection{Molecular structures, training data, feature definition, and causal relationships.} 

The benchmark dataset used to train and test the ML models was obtained from precise reactive molecular dynamics simulations of 1159 structures of carbon nanotubes, graphite,  defective and deformed graphitic assemblies, utilizing available experimental data and images to the extent available (Figure 2).\cite{jolowskyCNT2018_Liang}\cite{ColomerDWCNT2004} The simulations used the reactive interface force field (IFF-R), which quantitatively reproduces pi-pi stacking, surface and interfacial energies, Young's moduli and tensile strength of CNTs, graphene, and graphite in agreement with experiments.\cite{pramanik2017acsnano}\cite{winetrout2021iffr}  CNT morphologies in the training set included individual and bundled single-wall (Figure 2a), double-wall (Figure 2b inner wall highlighted in blue), and triple-wall CNTs (Figure 2c inner most wall highlighted in yellow). The nanotubes simulations were modeled with experimentally determined structural and defect considerations such as pristine (Figure 2a,d), with missing atoms (Figure 2b-d), with discontinuities or fracture defects (Figure 2d,f), and with reconfiguration or Stone-Wales defects (Figure 2d). Global features such as the number of nanotubes in a bundle, the type of nanotube modeled, the diameter of the nanotubes, the number of atoms and bonds in the simulation, and the cross-sectional area of the simulated bundle were identified as training features that could adequately capture the structural considerations that may influence CNT mechanical properties (Figure 2e). Tensile strength and tensile modulus are important parameters for designing safe materials for structural applications because they represent critical loads before material failure. The significance of strength and modulus for materials' design motivated the development of a machine learning model capable of predicting these two quantities for carbon nanotube bundles. The stress-strain curves of all structures were simulated up to failure to extract the tensile modulus and tensile strength (Figure 2f). Tensile strength is defined as the maximum stress before material failure. Tensile modulus is a metric that defines how hard a material is, and is calculated as the slope of the linear stress-strain response (typically 0.0-0.01 strain for hard materials). For a double-wall carbon nanotube (DWCNT) with fractured inner walls the tensile strength and tensile modulus were determined to be 19.2 GPa and 210 GPa respectively.

\begin{figure}[htbp]
    \centering
    \includegraphics[scale=0.32]{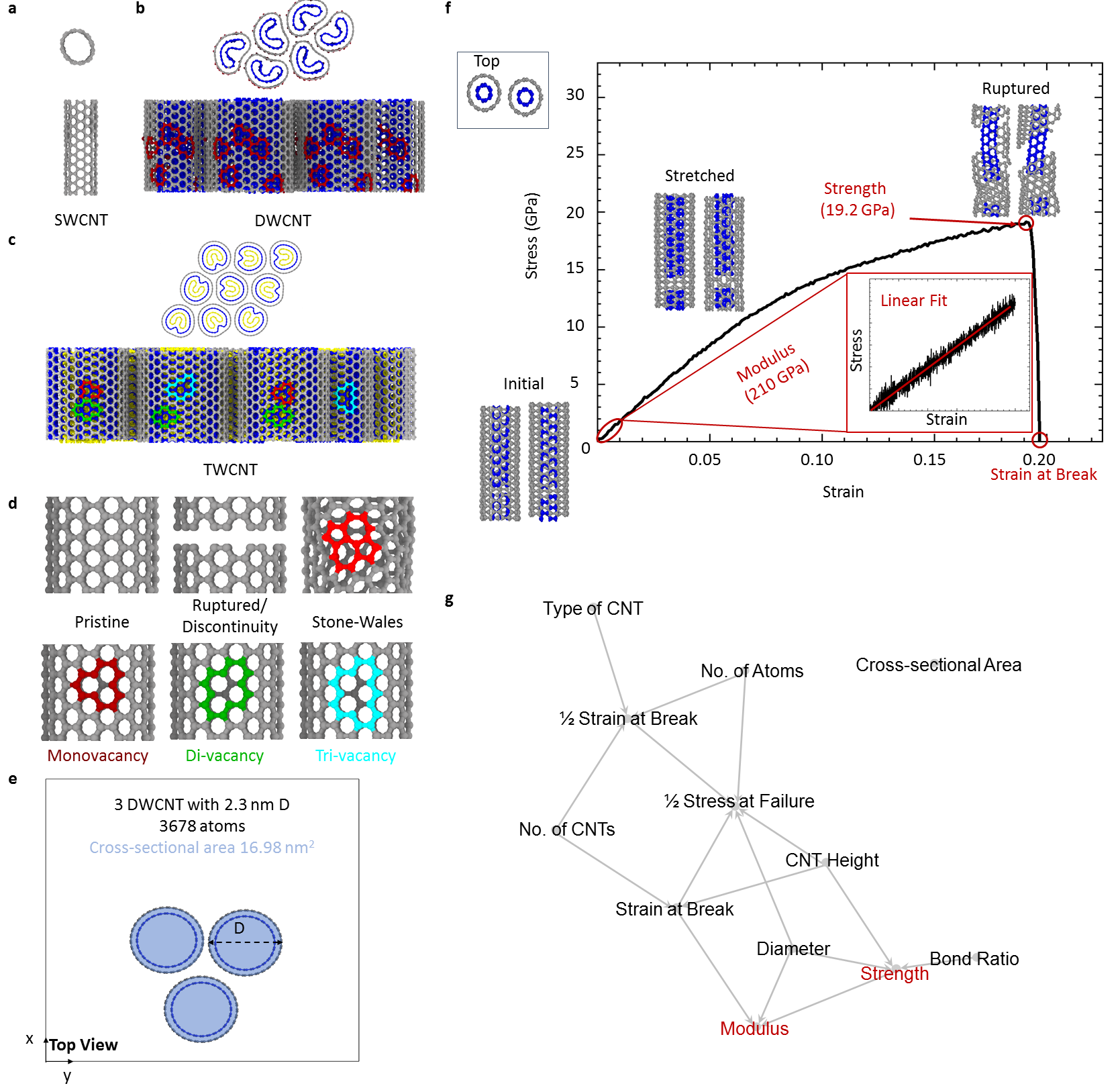}
    \caption{Overview of the molecular dynamics data set and methods that were used to train the hierarchical spatial graph neural network (\HGCN{}). Top and bottom view of (a) single-walled carbon nanotube with no defects, (b) bundle of double-walled carbon nanotubes with mono-vacancy defects, (c) dundle of triple-walled carbon nanotubes with mono-vacancy, di-vacancy, and tri-vacancy defects. (d) Overview of the structural features considered in the data set including defectless (pristine), discontinuous nanotubes, reconfiguration defects (Stone-Wales), and missing atoms (mono-, di-, and tri-vacancy).(e) Top-down view of a bundle of 3 DWCNT demonstrating the global features considered in the data set including the number of nanotubes in a bundle, the type (double-wall) of nanotube simulated, the diameter of each nanotube, number of atoms in the simulation, and cross-sectional area of the nanotube bundle. (f) Stress-strain curve from a tensile simulation of two double-walled carbon nanotubes with snapshots of the nanotubes at different strain points in the tensile simulation. Highlighted in red are most of the mechanical properties used for training \HGCN{} (strength, modulus, and strain at break). \HGCN{} was used to predict strength and modulus, so the values calculated from IFF-R are listed next to their labels in parentheses. (f) Relationship between some simulation features and their mechanical performance (strength and modulus highlighted in red), as determined by a causal additive model (CAM). The relationship determined by the CAM was used to inform the training of \HGCN{} in which the prediction of strength was used to predict modulus.
     }
    \label{Figure2}
\end{figure}

To rank the importance of each structural features' influence on CNT mechanical performance, structural feature significance for strength and modulus predictions were evaluated using a causal additive model (CAM), which aims to infer causal relations among features. \cite{buhlmann2014cam}  The CAM estimates strength as a direct cause of modulus (Figure 2g). The causal discovery information was added to our feature list and improvements by adding this relations into the ML predictions of modulus and strength will be discussed later. 

\clearpage

\subsection{Machine learning pipeline for property prediction}
\label{subsec:MLpipeline}

\begin{figure}[htbp]
    \centering
    \subfigure[]{\includegraphics[scale=0.45]{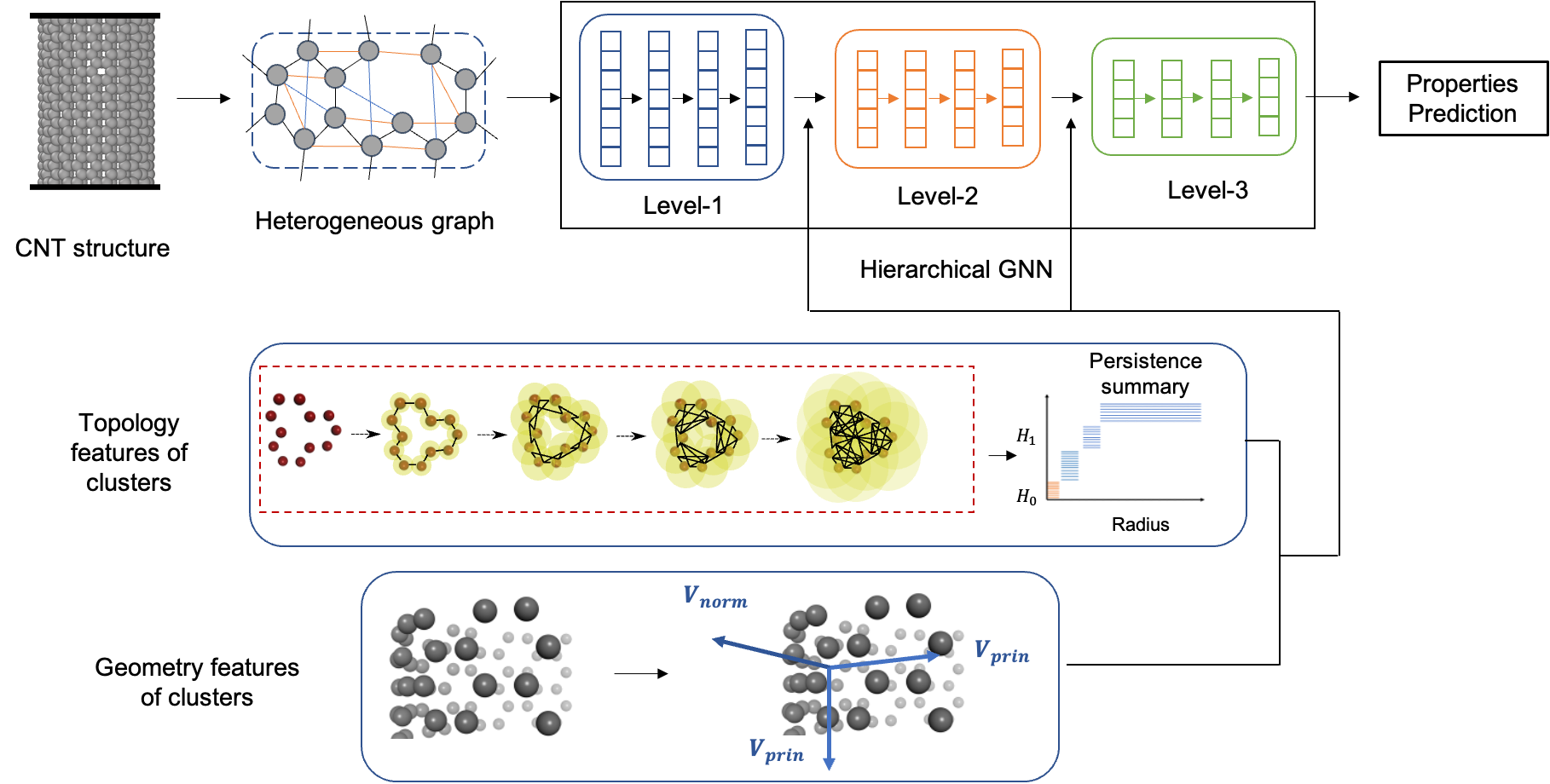}}
    \centering
    \subfigure[]{\includegraphics[scale=0.35]{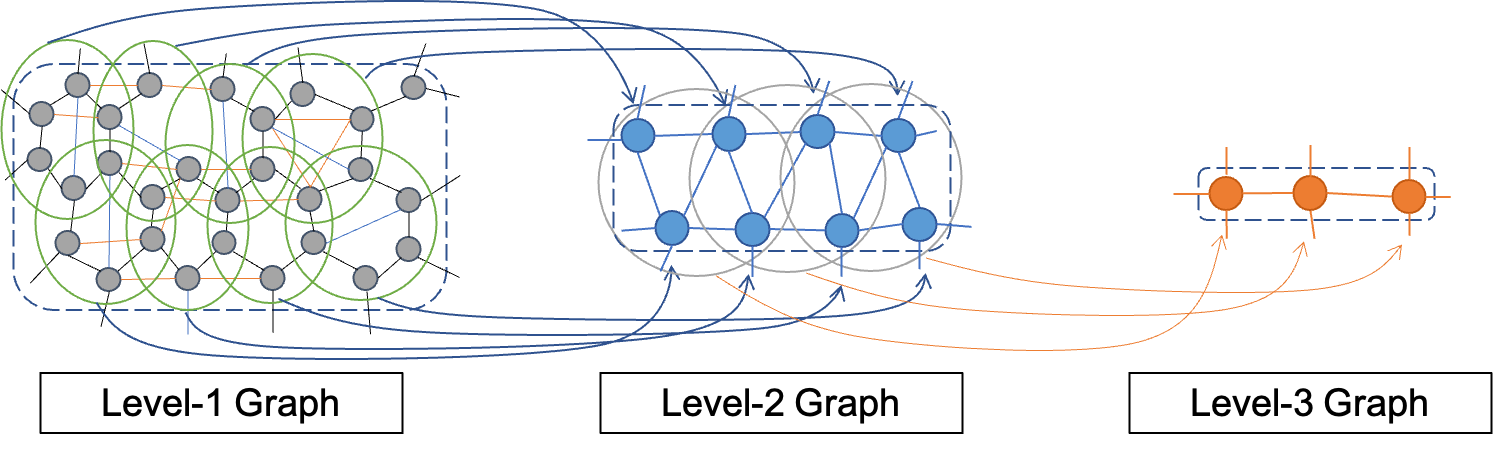}}
    \\
    \centering
    \subfigure[]{\includegraphics[scale=0.45]{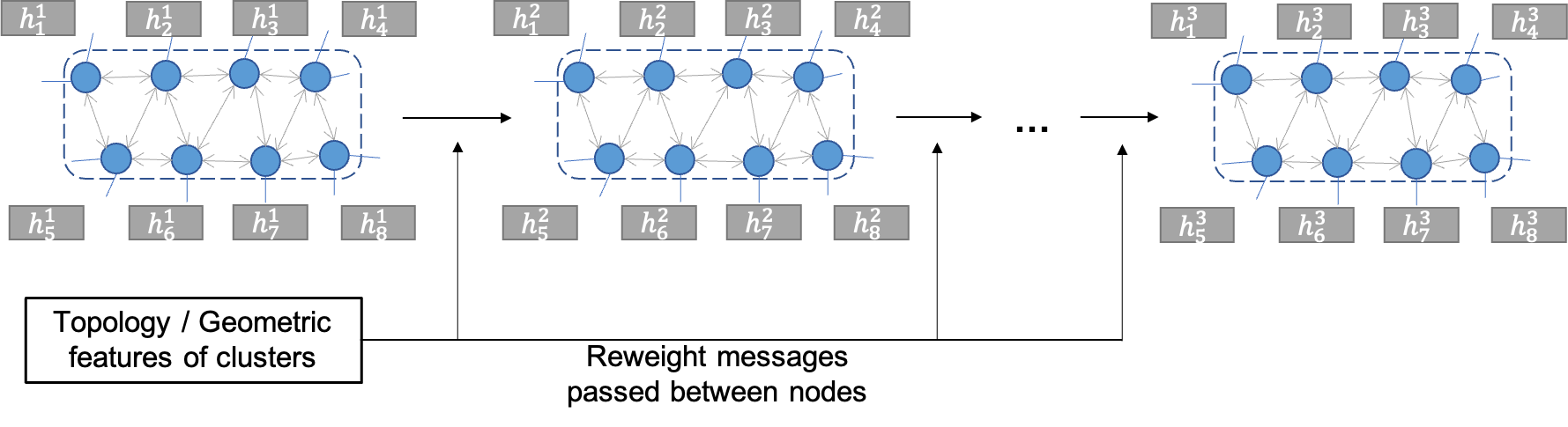}}
    \caption{The machine learning pipeline is a carefully designed hierarchical spatial GNN. (a) Overview of the pipeline to predict mechanical properties of CNTs and carbon nanostructures. Line 1 illustrates the generation of a heterogeneous graph for a given CNT structure and processing it with a hierarchical GNN. In each level of the hierarchy, we create a specific GNN to learn the graph representations and the list of vertical arrays within each level indicates the micro-layers. In addition, we compute topological and geometric features, more specifically, persistence summaries and PCA information of clusters from CNT structure. (b) Constructing the hierarchical graph series: In order to obtain a higher level graph, we group nodes in the lower level graph that are spatially close into clusters. Each node in the higher level graph represents a node cluster in the lower level, and we connect nodes if their corresponding clusters share overlap. (c) Illustration of the  message-passing workflow of the GNN in Level-2 and Level-3. Each node passes its features reweighted by geometric features of the corresponding cluster relative to the neighborhood.}
    \label{fig:MLpipeline}
\end{figure}

The high level framework of our machine learning pipeline, called \HGCN{} (Hierarchical Spatial Graph Neural Networks), for mechanical property predictions is shown in Figure \ref{fig:MLpipeline}. We describe the key ideas below. 
More details can be found in Section \ref{sec:hgcn}.

The input to \HGCN{} is the initial atomic structure $\atCNT$ of a CNT configuration. 
Instead of using a PointNet-type \cite{qi2017pointnet} architecture to take the set of atoms, viewed as a set of points each equipped with a radius, as input, we recognize the importance of both covalent bonds in the structure, as well as short-range interactions among atoms. Therefore we use a specially designed graph neural network-based architecture to process $\atCNT$ which we denote by \emph{\HGCN{}}. In particular, novel features of our \HGCN{} architecture include: a heterogeneous graph representation of input, a hierarchical neural network model so as to capture large-scale interaction among different parts of the CNT bundle, and the injection of local geometric and topological features to better encode the shape of CNT bundle into the neural network via attention mechanism. 
We briefly describe these key components below.

\begin{description}
   \item[Heterogeneous graph:] 
   It is common to use a \emph{bond-graph} $G_{bond\atCNT}$ to represent $\atCNT$, where each node corresponds to an atom, and there is an edge between two nodes if the corresponding atoms form a covalent bond. We go beyond the bond-graph and generate a heterogeneous graph $G_\atCNT$ with multiple types of edges among nodes from $\atCNT$. In particular, we add edges between nodes whose corresponding atoms are spatially close (i.e, atoms from different nanotubes but within 6\AA{} distance) or have small effective resistence distance. We also connect nodes whose corresponding atoms form a dihedral angle, the angle of two planes formed by four sequentially bonded atoms rotated about a central bond.
   This heterogeneous graph is then fed to a hierarchical GCN. 
   
   \item[Hierarchical GCN model:]
   A GCN (graph convolutional network) is a popular type of GNN (graph neural network; see Supplementary Material for a brief introduction). On the high level, a standard GCN takes a graph with initial node features as input. The input graph then goes through $\ell$ convolutional layers, during each of which it will update the feature vector stored at each graph node by aggregating features from neighboring nodes and performing a learned transformation of node features. See Figure \ref{fig:MLpipeline} (a, c) for an illustration. 
   However, it is known that GNNs tend to have the oversmoothing issue \cite{li2018deeper} and often cannot go very deep, thus limiting the aggregation of information from long range interactions. To address this issue, we use a hierarchical GCN 
   that process the input graph $G_\atCNT$ at multiple (three in our current experiments) resolutions: see Figure \ref{fig:MLpipeline} (a, b). Different from hierarchical GNN approaches proposed by \cite{ying2018hierarchical,huang2019attpool} etc., \HGCN{} finds hierarchies by spatial geometry information instead of graph topology or nodes features. The lowest level $L_1$ operates on input graph $G_\atCNT$; while in a higher level $L_i$, each node corresponds to a cluster of nodes of level $L_{i-1}$ and we can call such a node a \emph{super-node}. The feature vector associated to each super-node is obtained by a pooling layer on level $L_{i-1}$ graphs. 
   Within each resolution level $L_i$, we perform several GCN aggregation layers, which we refer to as micro-layers. 
    In the end, a last pooling layer is applied so as to obtain graph-level prediction.    
\item[Encoding local geometry/topology:] The shape of nanotubes and the spatial relation among neighboring tubes/sheets impact the final property. To encode local geometry, we use Principle Component Analysis (PCA) to capture a ``curvature"-like quantity for each resolution level $L_i (i> 1)$. 
To encode local interactions among spatially close atoms (from potentially different nanotubes), we use the so-called persistent homology (PH) to characterize the spatial distribution of points within each cluster in each level. Persistent homology is one of the most important developments in the field of topological data analysis in the past two decades and have already been applied to characterize different types of complex shapes \cite{edelsbrunner2000topological,edelsbrunner2010computational}. In our case, given a set of points (atom centers) forming a local cluster, by tracking the creation and death of topological features w.r.t. a growing sequence of space around these points, it can provide a meaningful yet succinct summary of the shape formed by these points. See Figure \ref{fig:MLpipeline} (d) for a simple illustration, and see Section {\bf Persistent Homology} in the Supplement for a brief introduction.
  
\item[Node features and edge attention:]
The above information are incorporated into our \HGCN{} via {\it node features} and via the \emph{edge attention mechanisms.} Given any graph node $v_a$ correspoinding to atom $a$, our initial node feature vector $\mu^{(0)}(v_a)$ in level $L_1$ includes the degree of $v_a$ in the bond graph (measuring local ``defect"), the 3D coordinates of atom $a$, as well as random generated features. For the higher level GNNs, the persistence summary of a cluster is included in the features of corresponding super-node. Node features will be transformed and aggregated through the neural networks. The edge-attention mechanism intuitively allows one to compute ``differential" of information at two endpoints $v_a, v_b$ of an edge ($v_a, v_b$) ($v_a$ and $v_b$ will be super-nodes/clusters in $L_j(j>1)$ level \HGCN{}), and use this to weight this connection (edge) $(v_a, v_b)$ when aggregating information for node $v_a$ from its neighbors. We use the two major principal vectors and the norm vector in local Principal Component Analysis (PCA) of points in each cluster for the edge-attention in $L_j$ level \HGCN. 
\end{description}

Figure \ref{fig:MLpipeline} illustrates the entire machine learning pipeline. Message passing is the scheme applied in GNN, while persistence summaries and geometric features are used in message passing process.

\clearpage

\subsection{Results on Modulus and Strength prediction}

Using the data set of 1159 tensile simulations of CNT bundles and graphitic structures, we trained \HGCN{} and evaluated the predictions of mechanical properties (Figure 4, 5).

\paragraph{Tensile properties distribution.} The distribution of tensile strengths within the training set is between 0-120 GPa, and the distribution of tensile moduli is between 0-1000 GPa. Few CNT bundles had strength and modulus values at the upper and lower limits attributing to strictly pristine or completely fractured cases respectively. However, even in highly crystalline CNT bundles it is impossible to achieve defect-less CNTs. To account for this, the training set is predominately comprised of CNT bundles that are flawed through missing atoms, Stone-Wales defects, discontinuities, or a combination of defects. The training set distribution reflects this, demonstrating a large portion of tensile strengths between 40-100 GPa and tensile moduli between 300-700 GPa (Figure 4 e,f). To achieve values of 0 GPa for tensile strength and modulus, some structure files generated after the tensile simulation were used as training structures. Completely fractured CNT bundles are not expected to be the primary CNT bundle morphology of interest, so a small population of completely fractured structures was included in the training set. 115 initial 3D computer generated CNT structures were used for testing of \HGCN{}. Test structures were randomly sampled to ensure that complex structural features such as straight and compressed nanotubes, pristine and defective structures, single and multi-wall nanotubes, etc. are captured in the test set.

\paragraph{Results.} The \HGCN{} method has shown to predict the tensile properties with a relative mean squared error (MSE) of 4.1\% deviation from the IFF-R simulations for strength, and 7.6\% deviation for tensile modulus when the strength prediction and pretrained \HGCN{} model (see Section \ref{sec:exp_setup} for details) is used to predict modulus (\HGCN{}-C) as determined by the CAM (Table \ref{tbl:res}). A 0.6\% MSE improvement is observed for predicting tensile modulus when using the strength prediction to predict modulus (\HGCN{}-B) is compared to \HGCN{} without using any property predictions as a input feature (\HGCN{}-A). 
In a more detailed analysis, tensile strength and modulus prediction error is dependent on tensile properties distribution in training set. (Figure 4). 
Tensile strength (Figure 4a) and tensile modulus (Figure 4b) prediction error was greater than or equal to a 20 GPa difference between the prediction and simulated values only for strength populations less than 3.7\% of the training set, for example, cases in which strength was simulated to be 0-40 GPa, or modulus between 0-200 GPa. The number of those cases in training set is relatively smaller as shown in Figure 4 e, f. Despite having a smaller portion in the training set, the predictions for CNT bundle cases close to the upper limit of performance (100-120 GPa and 800-1000 GPa for strength and modulus respectively) were below a 20 GPa difference between the prediction and simulated values, albeit with a larger prediction error deviation compared to the predominate sample range. The accurate predictions for a smaller portion in training set may be accredited to the less complex morphology of pristine CNT bundles that closely resemble somewhat flawed cases (Figure 2). On the other hand, fractured CNT bundles (0 GPa strength and modulus cases) can have dramatically different structures after material failure when compared to their unfractured counterparts (Figure 2f) leading to a higher property prediction error (Figure 4 a-d). What's more, \HGCN{} takes a server with 24 CPU cores and 1 RTX A6000 GPU about 7 hours to compute the persistence summaries and PCA information of CNT bundles in training set and train the model, but the inference process only takes about 2 minutes to obtain the predicted tensile strength and modulus for a CNT case in test set while IFF-R simulation takes 24 CPU cores about 30 minutes to compute the tensile properties.

\begin{figure}
    \centering
    \subfigure[]{\includegraphics[scale=0.42]{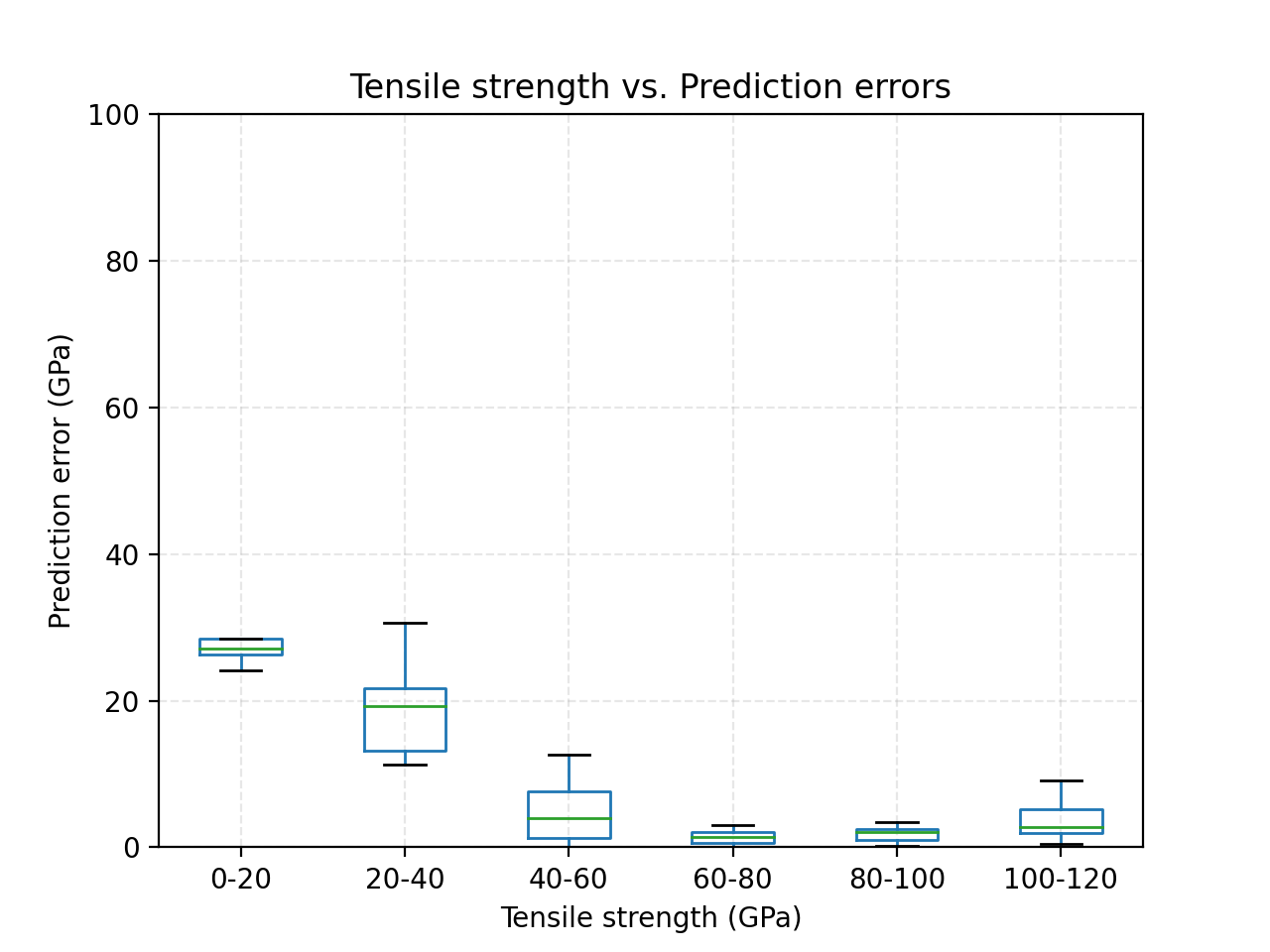}}
    \subfigure[]{\includegraphics[scale=0.42]{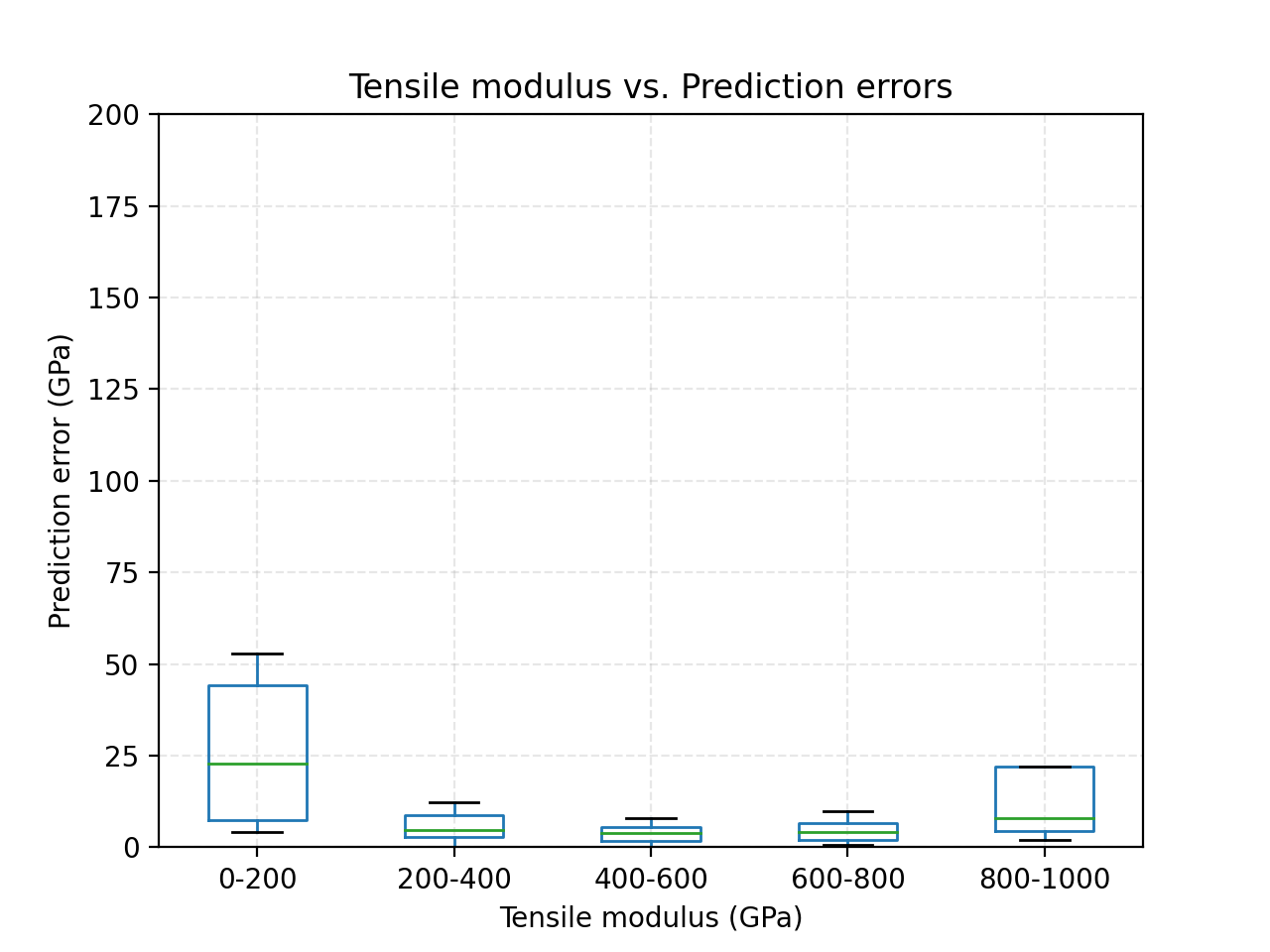}}
    \\
    \subfigure[]{\includegraphics[scale=0.42]{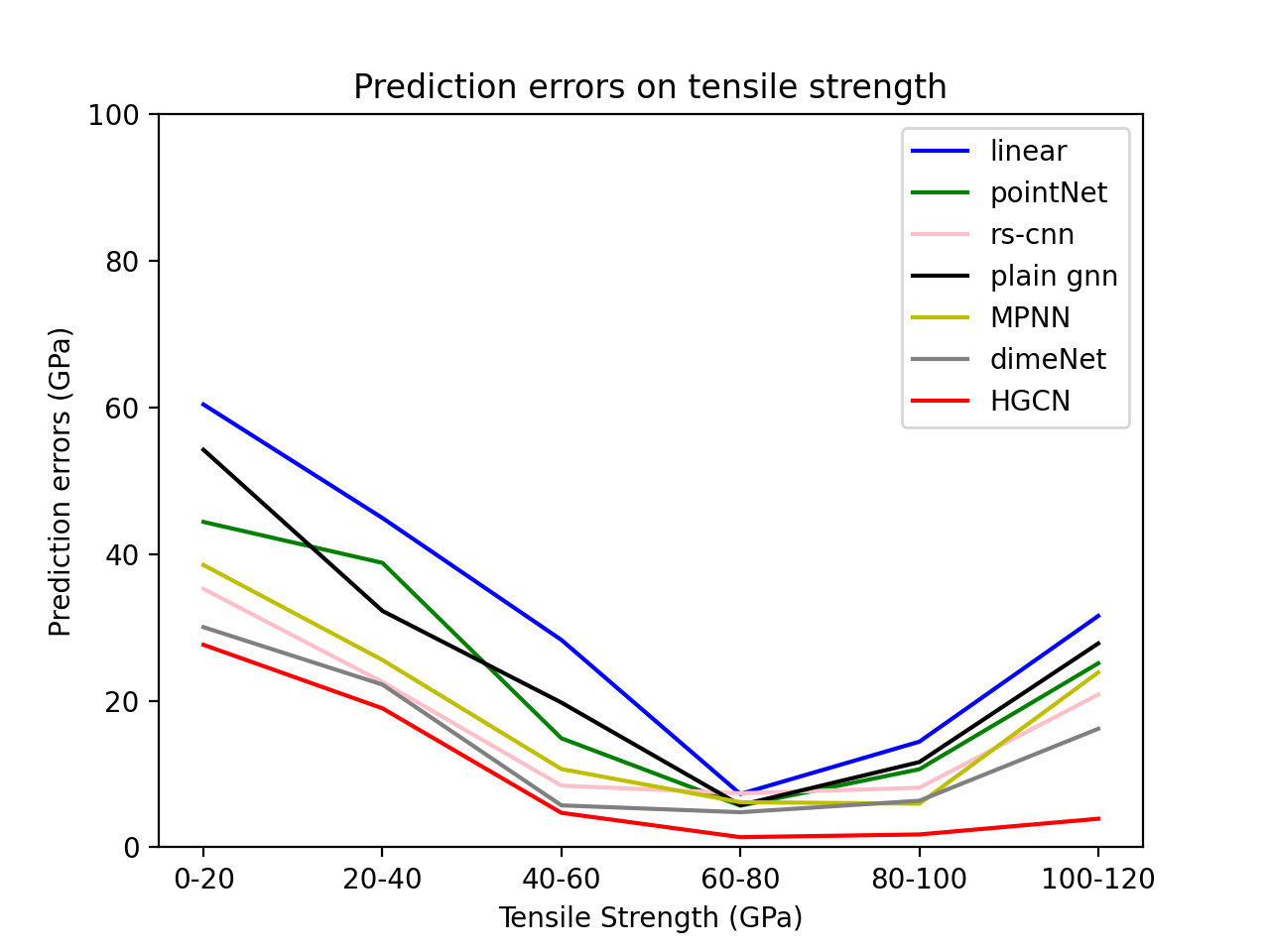}}
    \subfigure[]{\includegraphics[scale=0.42]{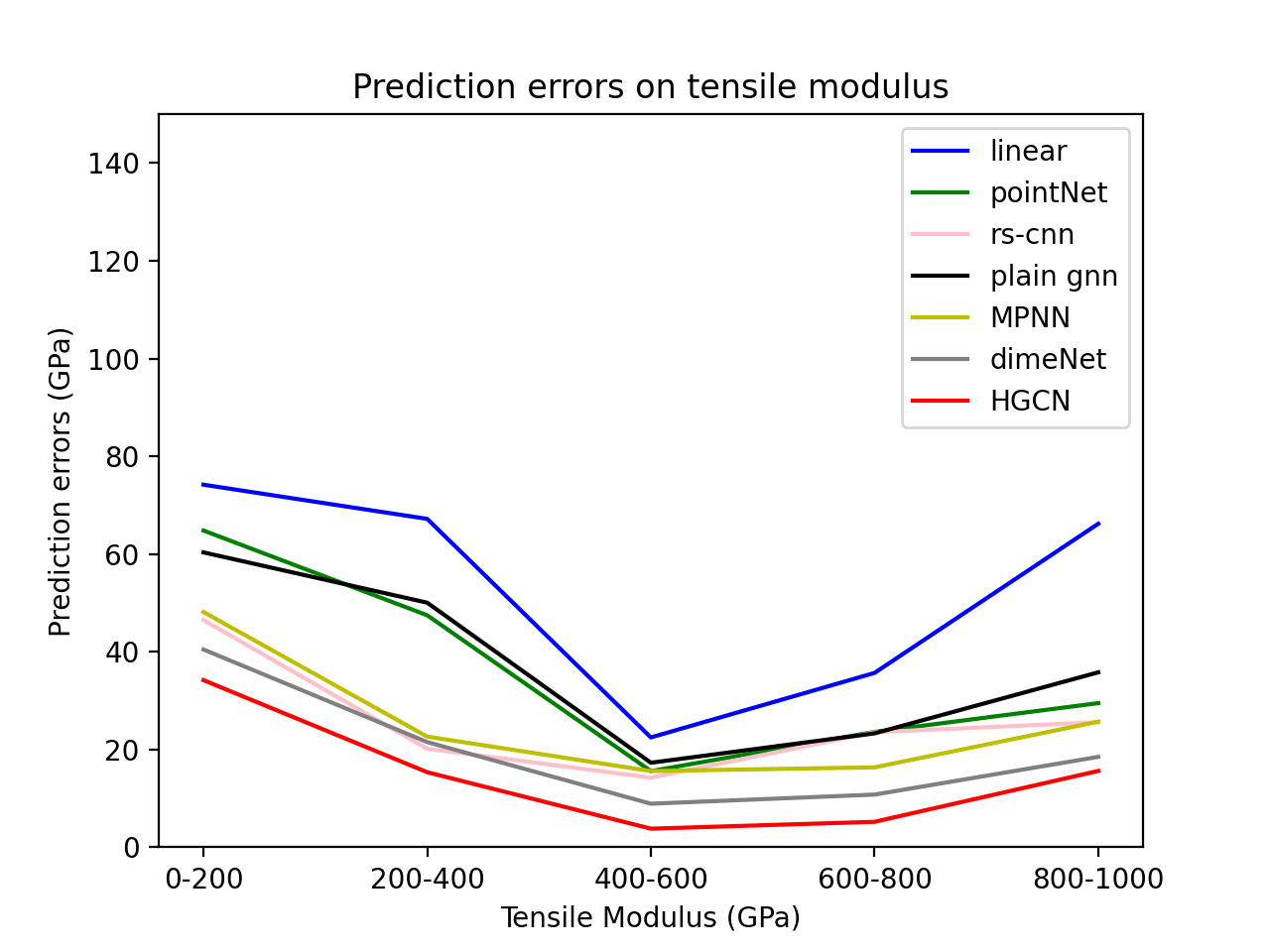}}
    \\
    \subfigure[]{\includegraphics[scale=0.42]{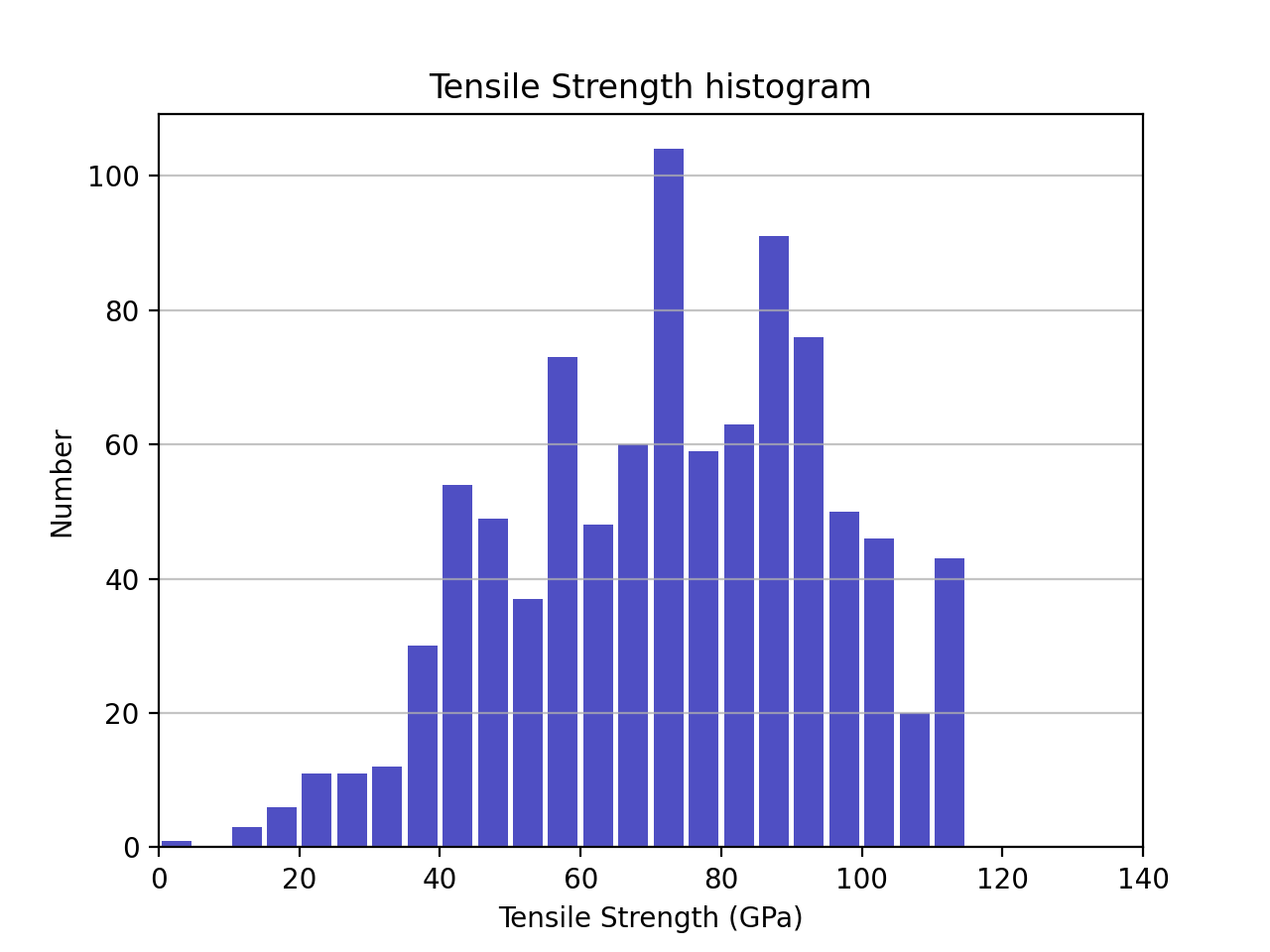}}
    \subfigure[]{\includegraphics[scale=0.42]{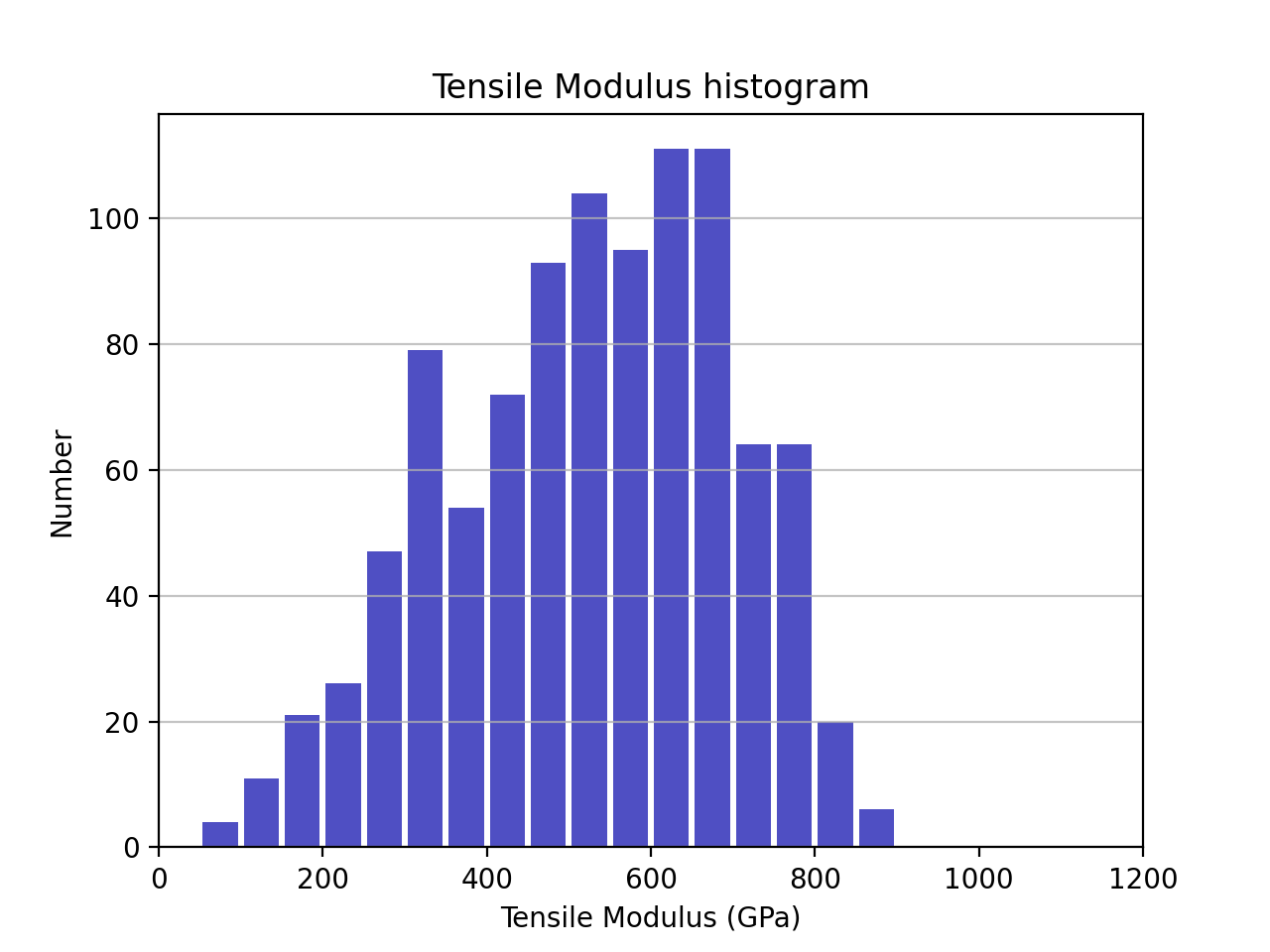}}
    \caption{Prediction of mechanical properties. (a) Box plot of prediction results of \HGCN{} on tensile strength. We divide the CNT-based models into 6 groups based on their tensile strength from 0 to 120 GPa. (b) Box plot of prediction results of \HGCN{} on tensile modulus. We divide the CNT-based models into 5 groups based on their tensile modulus from 0 to 1000 GPa. The Box plots show the average and standard deviation inside the box, as well as the maximum and minimum outside the box. (c, d) Comparison between \HGCN{} and 6 baselines on prediction over tensile strength and tensile modulus. The baselines are linear regression based on global features of CNT materials, PointNet and RS-CNN over point cloud formation CNT structures, GNN over bond graphs without the help of local topological and geometric features, MPNN and dimeNet taking simple geometric features like bond angles. (e,f) The distribution of tensile strengths and tensile moduli in the training dataset.
    }
    \label{fig:res}
\end{figure}

\paragraph{Comparison with other ML approaches.} 
We compare the performance of \HGCN{} to 6 baseline machine learning models, linear regression (LR) based on global features (i.e., number of nanotubes, number of atoms, diameter of a nanotube) of CNT bundles, PointNet \cite{qi2017pointnet} and RS-CNN \cite{liu2019relation} on point cloud formation CNT bundles, and three GNN models, vanilla-GNN \cite{kipf2016semi}, MPNN \cite{gilmer2017neural} and DimeNet \cite{klicpera2020directional}. Among them, DimeNet is considered to be the state-of-the-art in molecular structures properties prediction tasks. \HGCN{} outperforms all 6 baselines in predicting tensile strength and modulus of CNT bundles (Figure 4 c,d and Table \ref{tbl:res}). In general, all models follow the same prediction error trend by which a lower prediction error is observed for cases in the property ranges of more training cases (i.e. between 40-100 GPa for strength and 400-800 GPa for modulus). However, only while using \HGCN{} is a large upswing in prediction error not observed for the cases within the upper limit of strength (Figure 4c). For nearly all methods of training \HGCN{}, the mechanical property prediction accuracy for all non-large cases was 2-5 times better than baselines. 

\paragraph{Generalization to large structures.} 

\begin{figure}[ht]
    \centering
    \includegraphics[scale=0.45]{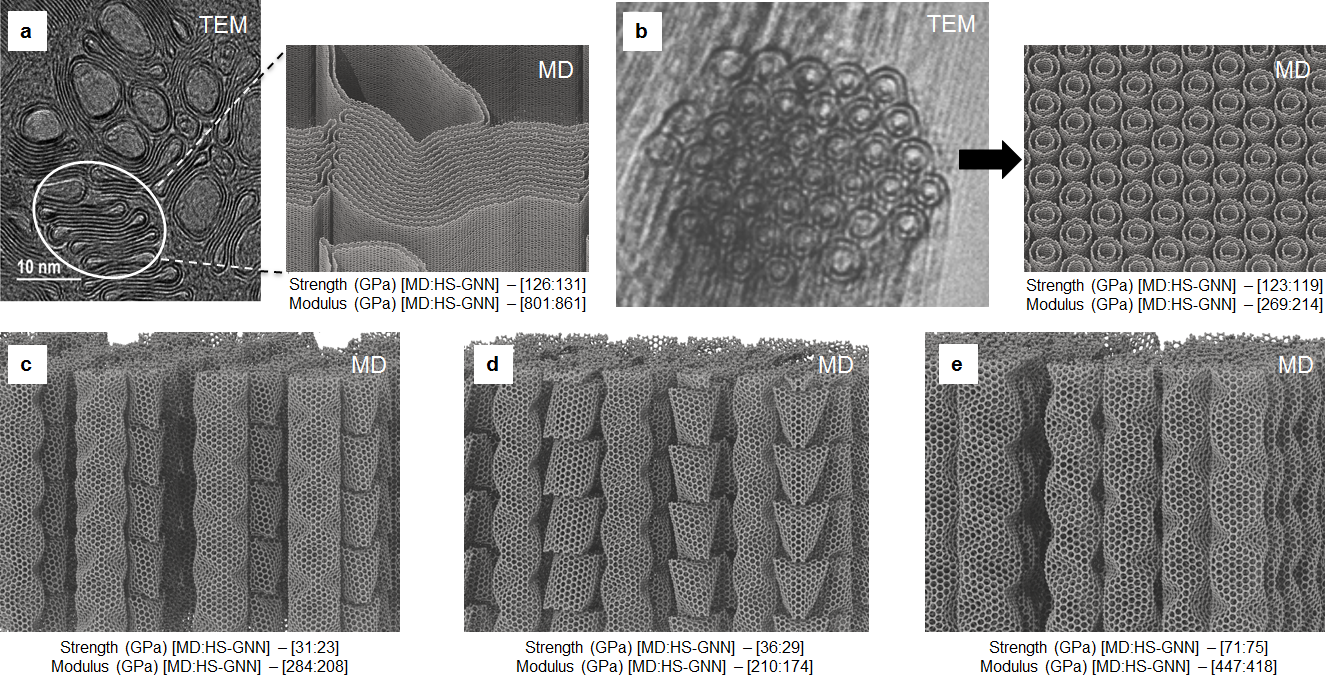}
    \caption{ 
    Comparison of 5 larger carbon nanostructures and calculated tensile strength and tensile modulus from molecular dynamics simulation (IFF-R) with the ML-prediction using the \HGCN{} method (under MD image). (a) Transmission electron microscopy (TEM) image of a CNT composite cross-section taken by Jolowsky et al. \cite{jolowskyCNT2018_Liang} (left) and corresponding models of flattened DWCNTs in MD simulation (right). The deviations are on the order of 5\%. (b) Hexagonal packing of a DWCNT bundle observed by TEM by Colomer et al. \cite{ColomerDWCNT2004} (left) and corresponding models in MD simulation (right). Differences between MD simulation and ML are small for strength and larger for modulus. (c-e) Defective structures derived from compression of SWCNT bundles. SWCNT bundle of sizes 72 (c) and 64 (d) nanotubes, each containing some nanotubes with fracture defects, and a SWCNT bundle of 64 nanotubes without defects. The model size is larger than the training set and uncertainties in ML predictions remain in the range of 4\% to 25\%.}
    \label{fig:large_cases}
\end{figure}

We also test our \HGCN{} on 16 larger (>20,000 atoms) CNT structures who consist of much more atoms than those in the training set, so as to see how well our trained model generalize as the size of input scales up. \HGCN{}-A method predicts strength with a MSE deviation of 12.7\% and modulus with a MSE deviation of 17.6\% from IFF-R simulations (Table 1). A 1.6\% MSE improvement is shown for \HGCN{}-B and 2.3\% MSE improvement for \HGCN{}-C, exemplifying the benefit of using property predictions as input features. For generalization of the trained ML models, \HGCN{} was nearly 3 times more accurate than vanilla-GNN or LR, and about 1.3 times more accurate than DimeNet (Table 1).
Figure \ref{fig:large_cases} highlights 5 of the 16 larger cases analyzed in-depth. It was determined that the tensile strength and tensile modulus predictions deviated less than  20\% from the IFF-R simulations in all larger cases except for the structures that were compressed and had fracture defects (Figure \ref{fig:large_cases} c, d), and for structures that had low densities and diameters much larger (i.e. 10 times larger diameters) than what was used in the training set. In Figure \ref{fig:large_cases}a, the MD structure is modelled after an experimentally determined CNT/polymer cross-section\cite{jolowskyCNT2018_Liang}, and has CNT diameters of 10-12 nm (about 10 times larger than in the training set). The accurate predictions (4.0\% strength deviation and 7.5\% modulus deviation) for Figure \ref{fig:large_cases}a can be attributed to the close packing of the CNTs, exemplified by the typical dog-bone formation of CNTs with diameters larger than 4 Angstroms\cite{elliot2004collapse}. Figure \ref{fig:large_cases}b demonstrates another simulation modeled after an experimentally determined cross-section of a double-wall CNT bundle with hexagonal packing and diameters of 1.45 nm\cite{ColomerDWCNT2004}. The structure in Figure \ref{fig:large_cases}b contains 64 CNTs in the bundle and 78080 atoms (about 6 times larger than the majority of the training set). However, \HGCN{} predicts a strength value for this structure with a value of about 3\% from the IFF-R prediction, suggesting that there is limited influence on the size of the input structure when predicting tensile strength for pristine CNT morphologies. 
When predicting the strength of large (>64 nanotubes), compressed CNT structures with fracture defects (Figure \ref{fig:large_cases} c,d) the strength predictions of \HGCN{} deviate 20\% or more. This suggests that complex morphologies with simultaneous fracture defects greatly diminish the performance of \HGCN{} more than any other observed factor. In contrast, when analyzing a similar morphology to Figures \ref{fig:large_cases}c,d without fracture defects (as shown in Figure \ref{fig:large_cases}e) it is shown that \HGCN{} is capable of predicting the tensile strength with less than 6\% deviation from IFF-R and tensile modulus with less than 6.5\% deviation from IFF-R. It should be mentioned, in the test set, cases 1051-1053 also had diameters of 12 nm. The strength and modulus deviations were >30\% for the cases 1051-1053, suggesting that CNT spacing in the input structure has some influence on prediction accuracy.
In summary, it is shown that the number of atoms is of little concern when generalizing to larger structures as long as the nanotube diameters are within the range of the training set (0.271-2.98 nm), and if there few, if any, fracture type defects.

\paragraph{Ablation study.} Besides the 3 GNN based baselines and our \HGCN{} model, we also evaluate two GNN setups. In the first setup, we found hierarchies within a CNT structure as \HGCN{}, but we didn't take topology and geometry information of each cluster as super-node features or edge attention. Its prediction error for test set on strength and modulus are 7.4\% and 14.6\%. In the second setup, we didn't design the hierarchical neural network anymore. But we computed persistence summaries and PCA information of neighborhood of each point, and took them as the input features of nodes or edge attention in GNN. Its prediction error on strength and modulus are 10.7\% and 19.5\%. Compared to performances of vanilla-GNN and \HGCN{}, we can conclude that both hierarchical design capturing long range interactions among atoms and local topological and geometric information contribute to our \HGCN{} model. The improvement from hierarchies plays a more important role than local topology and geometry.

\begin{table}[htbp]
    \centering
    \caption{Mean squared error in prediction of strength and modulus, using 90\% of the data for training. Mechanical property prediction error is also provided for larger cases (>20,000 atoms) not included in the data set and distinguished here by the proceeding (L). Our ML method was trained in 3 ways: \HGCN{}-A excludes using any mechanical property predictions as input features, \HGCN{}-B is trained using predicted strength to predict modulus, and \HGCN{}-C is trained using predicted strength and a pretrained model to predict modulus. Predictions using \HGCN{} methods are then compared to 6 baselines (right of the double vertical line). The \HGCN{} is 2 to 5 times more accurate even for larger structures not included in the training set. Uncertainties under 10\% can be considered competitive with experimental measurements, which often have similar errors. }
   
    \renewcommand\tabcolsep{4.5pt}
    \begin{tabular}{c|ccc||cccccc}
    \hline
    \footnotesize{Prediction error} & \footnotesize{\HGCN{}-A} & \footnotesize{\HGCN{}-B} & \footnotesize{\HGCN{}-C} & \footnotesize{LR} & \footnotesize{PointNet} & \footnotesize{RS-CNN} & \footnotesize{Vanilla-GNN} & \footnotesize{MPNN} & \footnotesize{DimeNet}\\
    \hline
    \footnotesize{Strength (\%)} & 4.1 & 4.1 & 4.1 & 21.3 & 15.18 & 11.60 & 14.5 & 12.02 & 8.6 \\
    \footnotesize{Modulus (\%)} & 8.8 & 8.2 & 7.6 & 37.4 & 30.01 & 20.55 & 31.2 & 19.72 & 16.4 \\
    \hline
    \footnotesize{Strength (L) (\%)} & 12.7 & 12.7 & 12.7 & 41.5 & 32.3 & 24.5 & 35.0 & 27.2 & 20.9\\
    \footnotesize{Modulus (L) (\%)} & 17.6 & 15.8 & 15.3 & 42.1 & 37.4 & 30.5 & 40.9 & 34.8 & 23.5 \\
    \hline
    \end{tabular}
    \label{tbl:res}
\end{table}

\clearpage

\section{Conclusion}
A hierarchical spatial graph neural network (\HGCN{}) has been developed and trained to predict the tensile strength and modulus of computer-generated models of pristine and defective carbon nanotube and graphitic assemblies. We used an extensible data set of 1159 unique molecular dynamics (MD) simulations of tensile stress-strain curves that were carried out using the IFF-R molecular dynamics force field, which accurately represents chemical bonding, physical properties, and bond dissociation. The average deviation using the \HGCN{} is 4\% for strength and 9\% for tensile modulus relative to the MD simulation, and by a factor of 2 to 5 lower than mechanical property predictions using other methods such as DimeNet, vanilla-GNN, PointNet, and linear regression. The incorporation of spatial information and use of a hierarchical method to process graph information is adequate for generalizing the neural network property predictions to larger graphitic structures outside the training set and still achieve less than 18\% deviation in ML-predicted mechanical property values relative to MD simulation. The ML model can be used for rapid screening of the mechanical performance of CNT morphologies and understanding the relationship between nanostructural features and carbon nanotube bundle performance for carbon fiber yarns used in automotive and aerospace applications. It is yet to be determined if the \HGCN{} can be generalized to accurately predict mechanical properties for nanostructures that are structurally similar but chemically different (e.g. boron nitride, macromolecular structures and composites), or structures that are chemically similar but structurally different (e.g. diamond nanofibers). 

\section{Methods}
\subsection{Building the Model Structure Database}
Models of carbon nanotube structures, assemblies, and defective graphitic structures were generated using the Materials Studio Graphical User Interface.\cite{MaterialsStudio2019} A total of 1159 different structures were created to build the structure database of CNTs and graphitic models. 
The CNT systems of interest include individual and bundled single-wall, double-wall, and triple-wall CNTs.  CNT bundles are defined as a group of more than one CNT. In general, the bundles ranged from 2 to 12 nanotubes in a simulation. Some larger structures were generated for generalizing our developed ML method, which contained up to 72 CNTs in a simulation. The diameter of the CNTs were modified by changing the chiral index values (N, M) using Materials Studio’s "Build Nanostructure" tool. For simplicity, all structures were generated such that N = M. CNTs with N and M values from 2 to 21 were evaluated.  The initial outer wall diameters of these CNTs ranged from 0.271 nm to 2.98 nm. In general, the height of the CNTs ranged from 2.42 nm to 2.52 nm. To introduce complex structural features, some CNT systems were compressed to a height of 2.29 nm before being strained in tension. The mechanical properties of pristine nanotubes and defective nanotubes were of interest. A defective nanotube is defined as a CNT that has one or more missing atoms such that some carbon atoms in the nanotube have 2 instead of 3 bonded neighbors, or the CNT lattice is structured such that pentagons and heptagons are formed instead of the typical hexagonal arrangement. Defects were implemented in three ways: firstly, by selectively severing bonds along the radial direction of the CNT such that nanotube is broken into two segments; secondly, by semi-randomly deleting atoms in the CNT structures effectively mimicking mono-, di-, and tri-vacancy defects commonly afflicting CNTs after synthesis, stress, or alignment; thirdly, by re-configuring the bonding of carbon atoms in the CNT lattice to form 2 heptagons and 2 pentagons, otherwise known as a Stone-Wales defect. In some cases, structures with 0 modulus and 0 strength were of interest, and these structures were obtained from the data file created after the CNT ruptured in tension during the simulation. IFF-R parameters were used to define the CNT atomic properties (supporting information S0). Each carbon atom's force field type was defined as cg1.

\subsection{Choice of Force Field}
In order to reproduce surface energies of graphitic materials that deviate less 5 percent from experimentally determined values, IFF incorporates virtual pi electrons on the corresponding carbon atoms. However, IFF is unable to simulate bond dissociation because of the use of a harmonic bond energy potential. To simulate bond dissociation IFF-R utilizes a Morse bond energy potential, which offers a better description for atomic bond behavior at large displacements. IFF-R maintains the use of virtual pi electrons and has been shown to reproduce experimentally determined surface energies and mechanical properties for graphite and CNTs, making it a suitable force field for investigating the influence of various nanostructural features on CNT mechanical performance (IFF-R reference). However, it should mentioned that virtual pi electrons were not considered for the current developed ML model for the sake of simplicity and ease of CNT structure generation.

\subsection{Reactive Molecular Dynamics Simulation to Create Training and Validation Data}
All structures were exported as .car/.mdf files so that they could be converted to a Large-scale Atomic/
Molecular Massively Parallel Simulator (LAMMPS) readable data file using the msi2lmp tool (part of the LAMMPS release).\cite{plimpton1995LAMMPS} The data file contains useful structural information such as the height of each CNT, the number of atoms in an individual CNT or a CNT bundle, and so on. The property predictions made from the developed machine learning pipeline used the data files as inputs. A data file was created for the initial CNT structures, equilibriated CNT structures, compressed CNT structures (where applicable), and the ruptured CNT  structures.  

A series of 1159 molecular dynamics simulations was carried out using the LAMMPS molecular dynamics simulation software. All simulations were run at 298 K using the canonical ensemble and Nose-Hoover thermostat and barostat. The Morse bond parameters were used for the carbon-carbon graphitic type bonds defined in IFF-R. The angles, dihedrals, and improper terms used class II potentials with a global cutoff of 12.0 Å. Each simulation was minim zed with an energy stopping criteria of 1e-4 and a force stopping criteria of 1e-6 kcal/molÅ. Before the tensile simulation began each simulation was allowed to equilibrate for 10000 femtoseconds. When applicable, the simulation cell was reduced in size after equilibration for 3500 femtoseconds and at a rate of -20/ns (7\% of the original CNT height) to form the compressed structures. The tensile simulation was run until failure (no more than 100,000 femtoseconds were needed) and at an engineering strain rate of 20/ns.

\subsection{Generation of Features}
\subsection{Molecular Dynamics Mechanical Property Calculations}
    The mechanical properties were calculated using Microsoft Excel and an example can be found in the supporting information. The stress, strength, and Young's Modulus were calculated in units of GPa. To calculate stress the pressure tensor along the axial direction of the CNT (zz-direction) was normalized to account for only the stress experienced by the CNTs. The normalization involves computing the total cross-sectional area of the CNTs by using \[A = N\pi(r+0.19)^2\]where N is the number of CNTs and r is the CNT radius plus a van der Waals correction of 0.19. The total area of the simulation cell is then divided by the normalized area (A) to get the correction term, C. Stress can then be defined as \[\sigma = |C\Delta P_{zz}| \;\;\; where \;\;\; \Delta P_{zz} = P_t - P_0\] 
    here the subscript t is the value at any time step and the subscript 0 is the initial value after equilibration and compression. Strain is calculated by \[\epsilon = \frac{\Delta L_{zz}}{L_0}\;\;\; where \;\;\; \Delta L_{zz} = L_t - L_0\]
    The strength was determined to be the maximum corrected stress value and modulus was calculated by taking the slope of the stress-strain curve where the response is linear (e.g. between 0-0.01 strain). For the compressed structures, the Young's Modulus was calculated after the CNTs were stretched to their original length (e.g. after 0.07 strain).

\subsection{\HGCN{} Neural network architecture}
\label{sec:hgcn}
The overall \HGCN{} pipeline is given in Figure \ref{fig:MLpipeline} with a brief description in Section \ref{subsec:MLpipeline}. We now describe the components in more details. 

\subsubsection{Hierarchical heterogeneous graph formation}

\paragraph{Heterogeneous graph representation of CNT structures.}A CNT structure can be modeled as a graph in different ways. The bond graph $G_{bond\atCNT}$ captures the information of chemical bond connections which plays a pivot role in molecular structures. However, it ignores the spatial relations between nearby but non-bonded atoms (e.g, two nearby carbon atoms from neighboring nanotubes). This issue can be addressed by connecting nearby pairs of atoms, namely, two graph nodes are connected if their corresponding atoms are within certain Euclidean distance $c_e$ to each other. In addition, effective resistance distance \cite{babic2002resistance} between two nodes in a graph (network) can measure how well the two nodes are connected via paths in the graph, which can reflect the graph topology. Thus we also connect pairs of nodes if their effective resistance distance in the bound graph $G_{bond\atCNT}$ is smaller than a threshold $c_r$. 
Finally, we also connect nodes whose corresponding atoms form a dihedral angle. All these together give rise to a \emph{heterogeneous graph} $G_\atCNT = (V, E)$ with four different types of edges $E = E_b \cup E_e \cup E_r \cup E_d$,  representing the edge sets formed by chemical bonds, Euclidean distance,  effective resistance distance, and dihedrals, respectively. 

\paragraph{Creating hierarchical representations.} Existing GNN models suffer the so-called over-smoothing issue, where signals on graph are quickly smoothed out as the number of layers increase. Thus a GNN usually has only a small number of layers, causing the receptive field of each node limited to a relatively local region when a GNN model is applied on large size graphs. We design a hierarchical architecture leveraging spatial geometric information to tackle this issue. 

An input CNT $\atCNT$ can be viewed as a point cloud $V_\atCNT$ with each point representing the center of an atom in $\atCNT$. 
We first apply a hierarchical clustering algorithm on this 3D point cloud $V_\atCNT$, and obtain a series of coarser and coarser point sets $V^{(1)} = V_\atCNT \subset V^{(2)} \subset  \ldots \subset V^{(k)}$. In particular, $V^{(i+1)}$ is obtained by taking a so-called $\delta_i$-net of point set $V^{(i)}$, for each $i\in [1, k)$. 
The computation of a $\delta$-net $Q$ of a point set $V$ is given in Algorithm \ref{alg:clustering} in Supplement C.
Note that each point (called a \emph{super-node}) in $V^{(i+1)}$ corresponds to a cluster of points in $V^{(i)}$. 

 The coarser point set $V^{(i)}$ will serve as the node set for a coarser graph $G^{(i)}$ in level $L_i$ for any $i \in [1,k]$. Set $G^{(1)} = G_\atCNT$ as the graph representation in level $L_1$. 
 For any $i > 1$, we connect two super-nodes in $V^{(i)}$ if there exists overlapping nodes from their corresponding clusters (recall each cluster consists of a set of nodes in $V^{(i-1)}$) in $G^{(i-1)}$.  

\subsubsection{Heterogeneous GNN in $L_1$ level hierarchy}
The input to the GNN in level $L_1$ is the heterogeneous graphs $G_\atCNT$ with 4 types of edges $E = E_b \cup E_e \cup E_r \cup E_d$. 
The GNN used in level $L_1$ consists of $q$ GIN layers \cite{xu2018powerful}, followed by $p$ GAT layers \cite{velivckovic2017graph}, and it will train different weights for different types of edges. (See Supplement B for the message passing/aggregation in the GIN or GAT layers.) 
We now describe this in details using the edge set $E_b$ as an example: 

Suppose there are $N_\atCNT$ nodes in $G_\atCNT$ ($|V| = N_\atCNT$). We denote the node (feature) representation matrix in the $l$-th ($l=1,2, ...,q+p$) layer as $H_l=[h_1^{(l)}, h_2^{(l)},...,h_{N_\atCNT}^{(l)}]^T$ in which $h^{(l)}_i$ is feature representation of the $i$-th node in the $\ell$-th layer. The input node features are denoted as $H_0 = [h_1^{(0)}, \ldots, h_{N_\atCNT}^{(0)}]^T$, where each $h_i^{(0)}$ consists of the $i$-th atom's 3D coordinates, degree, and randomly generated features.

For $l=1,...,q$, $G_\atCNT$ is processed by GIN layers (Equation (\ref{eq:gin}) in Supplement), and we process $G_\atCNT$ (where nodes are equipped with feature representations $H_q$ output from the $q$th GIN layer) by $p$ GAT layer for $l=q+1,...,q+p$. From Equation (\ref{eq:gat}) in Supplement we obtain the final node representations $H_b=[h_{1;b}, h_{2;b}, ..., h_{N_s;b}]^T$ according to chemical bond edge set $E_b$.

Using the same GNN architectures as described in Equations (\ref{eq:gin}) and (\ref{eq:gat}) on edge sets $E_d$, $E_e$ and $E_r$, respectively, we have node representations $H_{*} = [h_{1;*}, h_{2;*}, ..., h_{N_s;*}]^T$ (* denotes $d, e, r$). We get final node representations $H=[h_1, h_2, ..., h_{N_s}]^T$ by integrating the 4 kinds representations into the convolution layer as follows:
\begin{equation}\label{eqn:finalnodes}
    \bar{h}^{(1)}_u = \mathrm{ReLU}(W'(h_{u;b} || h_{u;d} || h_{u;e} || h_{u;r}))
\end{equation}
where $\cdot||\cdot$ stands for concatenation.

\subsubsection{Spatial information enhanced GNN in higher level hierarchies (i.e., $L_i$ for $i > 1$)}

Level $L_1$ outputs node feature representations $\bar{h}^{(1)}_u$ as in Eqn (\ref{eqn:finalnodes}). 
In general, suppose we have already finished processing level $L_{k-1}$ with the final node representations $\bar{h}^{(k-1)}_u$ for nodes in $u\in V_{k-1}$ in graph $G^{(k-1)}$. 
In level $L_k$, the graph we will process is the coarser graph $G^{(k)} = (V_k, E_k)$. 
In order to initialize node features for nodes in $V_k \subseteq V_{k-1}$, recall that each node $v \in V_k$ in fact corresponds to a cluster of nodes $C_v \subseteq V_{k-1}$. 
Let $G^{(k-1)}(v)$ denote the subgraph of $G^{(k-1)}$ spanned by nodes in $C_v$ -- intuitively, this subgraph $G^{(k-1)}(v)$ from graph $G^{(k-1})$ in level $L_{k-1}$ is collapsed into a single node $v$ in graph $G^{(k)}$ in level $L_k$. We simply perform a max-pooling (see Equation (\ref{eq:max-pooling})) of the node features of subgraph $G^{(k-1)}(v)$ to obtain a feature representation $h_{G^{(k-1)}_v}$ for the entire subgraph $G^{(k-1)}(v)$.
\begin{equation}
\label{eq:max-pooling}
    h_{G^{(k-1)}_v} = [\mathrm{max}(\{\bar{h}^{(k-1)}_u[0]|u \in V_{k-1}\}), \mathrm{max}(\{\bar{h}^{(k-1)}_u[1]|u \in V_{k-1}\}),\cdots, \mathrm{max}(\{\bar{h}^{(k-1)}_u[d_{k-1}]|u \in V_{k-1}\})]
\end{equation}
where $d_{k-1}$ is dimension of $\bar{h}^{(k-1)}_u$. This representation $h^{(k), 0}_v = h_{G^{(k-1)}_v}$ is then used as the initial feature for node $v\in V_k$. 

Now we have the coarse graph $G^{(k)}$ with initial features $H^{(k), 0} = [h^{(k), 0}_1, \ldots, h^{(k), 0}_{N_k}]^T$ for all $N_k = |V_k|$ nodes in $V_k$. The GNN for level $L_k$ consists of $r$ GCN layers \cite{kipf2016semi} as GCN layer is easier to add reweighting factors introduced in the following than GIN layer.
More precisely, in the $l$-th layer ($l = 1, ..., r$) in hierarchical level $L_k$, the message passing function a node $u\in V_k$ is: 
\begin{equation}
    m_u^{(l)} = \sigma(\sum_{v \in u} W^{(l)} h_v^{(l-1)})
\end{equation}
where $\sigma(\cdot)$ is a non-linear function like $\mathrm{ReLU}(\cdot)$, and the parameter $W^{(l)}$ is a linear transformation matrix that will be learned by training. 

In our model, we add a reweighting factor $\tau_{uv}$ for the message between two nodes, $u$ and $v$, during graph convolution: 
\begin{equation}
\label{eq:re_gcn}
    m_u^{(l)} = \sigma(\sum_{v \in N(u)} \tau_{uv}^{(l)} W^{(l)} h_v^{(l-1)})
\end{equation}
This reweighting factor can tell the difference and significance of any message based on local topological and geometric features at nodes, or in the corresponding subgraphs. We describe a method to learn this factor in Supplement C.
After the message aggregation, we take a 3 layer MLP as the update function in a GCN.
\begin{equation}
    h_u^{(l)} = \mathbf{MLP}(h_u^{(l-1)}, m_u^{(l)})
\end{equation}

After 3 layers convolution, we could obtain a graph representation for any {\bf subgraph $G'$ of $G^{(k)}$} based on the node representation of nodes in $G'$ through max-pooling the same as what used before for level $L_1$. 

In the highest level hierarchy, we add one regression layer predicting the tensile properties taking both graph representations and the global geometric and topological features as input in the final step. Specifically, we take the global persistence summaries (see Supplement A) of all the atoms in the CNT bundles in the final regression layer.

\subsection{Property prediction}
\label{sec:exp_setup}

\begin{description}
    \item [Dataset:]  Our dataset has 1159 CNT bundles which we split into training and test dataset as the ratio of 9:1. Each CNT bundle consists of one or more carbon nanotubes, and preserves 3D positions of all carbon atoms ($(x,y,z)$ coordinates), chemical bonds between atoms as well as chemical dihedrals among atoms. The sizes of CNT bundles vary from 480 atoms to over 10,000 atoms. We denote those features a CNT bundle as $(V, E_{b}, E_{d})$, where $V=\{(x_i, y_i, z_i) | i = 1, \cdots, N\}$ is the set of atoms, $E_b$ is the set of chemical bonds and $E_d$ is the set of chemical dihedral relations.
    Our task is to predict two mechanism properties, tensile strength and tensile modulus, based on those input features.
    
    \item[Architecture:] We take a 3 level hierarchical \HGCN{}. As we introduced in Section (\ref{sec:hgcn}), we take $\delta$-nets to get the coarser graphs. In particular, we take a $\delta_1$-Net to obtain the level $L_2$ graph $G^{(2)}$ from the original level $L_1$ graph $G^{(1)} = G_\atCNT$, and we take  a $\delta_2$-Net to obtain the level $L_3$ graph $G^{(3)}$ from $G^{(2)}$. 
To choose the parameter $\delta_1$ and $\delta_2$, we randomly sample 100 CNT bundles from training set, compute the average 3D Euclidean distance (denoted by $d_E$) between every two nodes connected by chemical bonds, and set $\delta_1 = 15 d_E$ and $\delta_2 = 30 d_E$.
    
    We then design GNN architecture for each level. In the basic $L_1$ level, a heterogeneous graph $G^{(1)}=(V, E_b, E_d, E_e, E_r)$ is formulated from each point set $V$. If there exists a chemical bond between points $u$ and $v$, we add an edge $(u, v)_b$ in $E_b$. 
    If there two points $u$ and $v$ can formulate a chemical dihedral angle with a third point, we add an edge $(u,v)_d$ in $E_d$. If Euclidean distance between two points $u$ and $v$ is smaller than a cut-off distance $c_E = 6 d_E$, we add an edge $(u, v)_e$ in $E_{e}$. In the final, we compute the average effective resistance distance between every two nodes connected by chemical bonds in 100 sampled CNT bundles and denote it as $d_r$. We add an edge $(u, v)_r$ in $E_r$ if effective resistance distance between two points $u$ and $v$ are smaller than $c_r = 10 d_r$. For each relation in $\{E_b, E_d, E_e, E_r\}$, we take $q$ GIN layers followed by $p$ GAT layers to process as we mentioned in Section (\ref{sec:hgcn}). In each higher level, a coarser graph is processed through $r$ reweighted GCN layers. Thus we have 3 hyperparameters, $q$, $p$ and $r$, to tune. They are chosen from $\{3,4,5,6\}$, $\{1,2,3\}$, and $\{3,4,5,6\}$ respectively. We tune those hyperparamters by 5 fold cross-validation. In a more explicit manner, we split the training set into 5 folds, each time we use 4 folds to train our model, and then compute the prediction error on the rest 1 fold as validation set. Repeat this operation 5 times so that each fold is taken as validation set once. We take the average mean squared error over 5 iterations as the evaluation performance of a model with certain hyperparameters. Finally we test the model and hyperparameters with best evaluation performance on our test set and report the results.
    
    As an exploratory study, we predict tensile strength and tensile modulus in two ways. The first way is to apply generalized additive models \cite{wood2017generalized} to predict tensile strength and tensile modulus using only features (e.g., CNT diameter, CNT height) separately. The second way is to  first estimate causal relations among tensile strength, tensile modulus, and features using a causal discovery method, CAM \cite{buhlmann2014cam}, and then utilize the inferred causal relations between tensile strength and tensile modulus in prediction. As shown in Figure (\ref{Figure2} g), the CAM finds that tensile strength is a direct cause of tensile modulus. When we only use features to predict tensile modulus, the generalized additive model yields a fitted $R^2$=0.931 (the higher, the better). When we add tensile strength in predicting tensile modulus besides features, the $R^2$ increases to 0.941. This exploration  suggests that using tensile strength helps the prediction of tensile modulus. We therefore design three different final prediction setups:
(A) We train two models to predict tensile strength and tensile modulus independently. (B) We first train a model to predict tensile strength. We then train a second model to predict tensile modulus which takes predicted strength as part of input features in the final layer. 
(C) We train a model to predict tensile strength first. We then take the trained GNN parameters as initialization to train a second model to predict tensile modulus which also takes the predicted strength as input features in the final layer to predict modulus. 

\end{description}

\section*{Acknowledgments}
The authors acknowledge support by the National Science Foundation (OAC-1931587, CMMI-1940335) and the University of Colorado Boulder. Work by Wang and Zhao are partially supported by NSF under grants OAC-2039794 and CCF-2051197. The allocation of computational resources is acknowledged at the Argonne Leadership Computing Facility, which is a DOE Office of Science User Facility supported
under contract DE-AC02-06CH11357, and at the Summit supercomputer supported by the National Science Foundation
(ACI-1532235 and ACI-1532236).

\bibliographystyle{unsrt}
\bibliography{reference}

\appendix 

\section{Persistent Homology}
\label{sec:ph}
Persistent homology \cite{edelsbrunner2000topological,edelsbrunner2010computational} is one of the most important developments in the field of topological data analysis, and persistence summaries are often used as features in statistics or machine learning tasks associated with graphs, point clouds and 3-D shapes \cite{bubenik2015statistical,Kusano2017Kernel,Adams2017Persistence,Carri2017Sliced,zhao2019learning}. Suppose we are given a topological space $X$ and a $filtration$, a sequence of growing subsets, of $X$: $X_1 \subseteq X_2 \cdots \subseteq X_n = X$. As we inspect $X$ through this filtration, sometimes a new topological feature like a void or a loop is created when entering $X_i$, and destroyed in $X_j$. Persistent homology can capture the birth and death of topological features in the form of a $persistence$ $diagram$ $\dgm X$. Specifically, the $k$-dimensional persistence diagram $\dgm_k X$ consists of a multi-set of persistence points in the birth-death plane. Each persistence point $(b, d)$ indicates that a $k$-dimensional topological feature appears when entering $X_b$ and disappears upon entering $X_d$. The $persistence$ of this feature is its lifespan $|d-b|$.
In other words, the persistence diagram provides a simple yet rich summary for the entire evolution of space $X$ through the lens of the filtration $X_1 \subseteq X_2 \cdots X_n=X$, encoding multi-scale features in $X$. 

When $X$ is a point cloud, one common approach to obtain persistent summaries is constructing a so-called \emph{Vietoris-Rips filtration}. Here, a space is modeled by a \emph{simplicial complex} spanned by a vertex set $V$: Roughly speaking, a $k$-dimensional simplex is the $k$-dimensional generalization of vertices ($0$-D), edges ($1$-D) and triangles ($2$-D simplices). A simplicial complex is then simply a union of simplices with the condition that if a simplex is contained in this complex, then any of its face will also be in the complex. 

\begin{figure}
    \centering
    \includegraphics[scale=0.5]{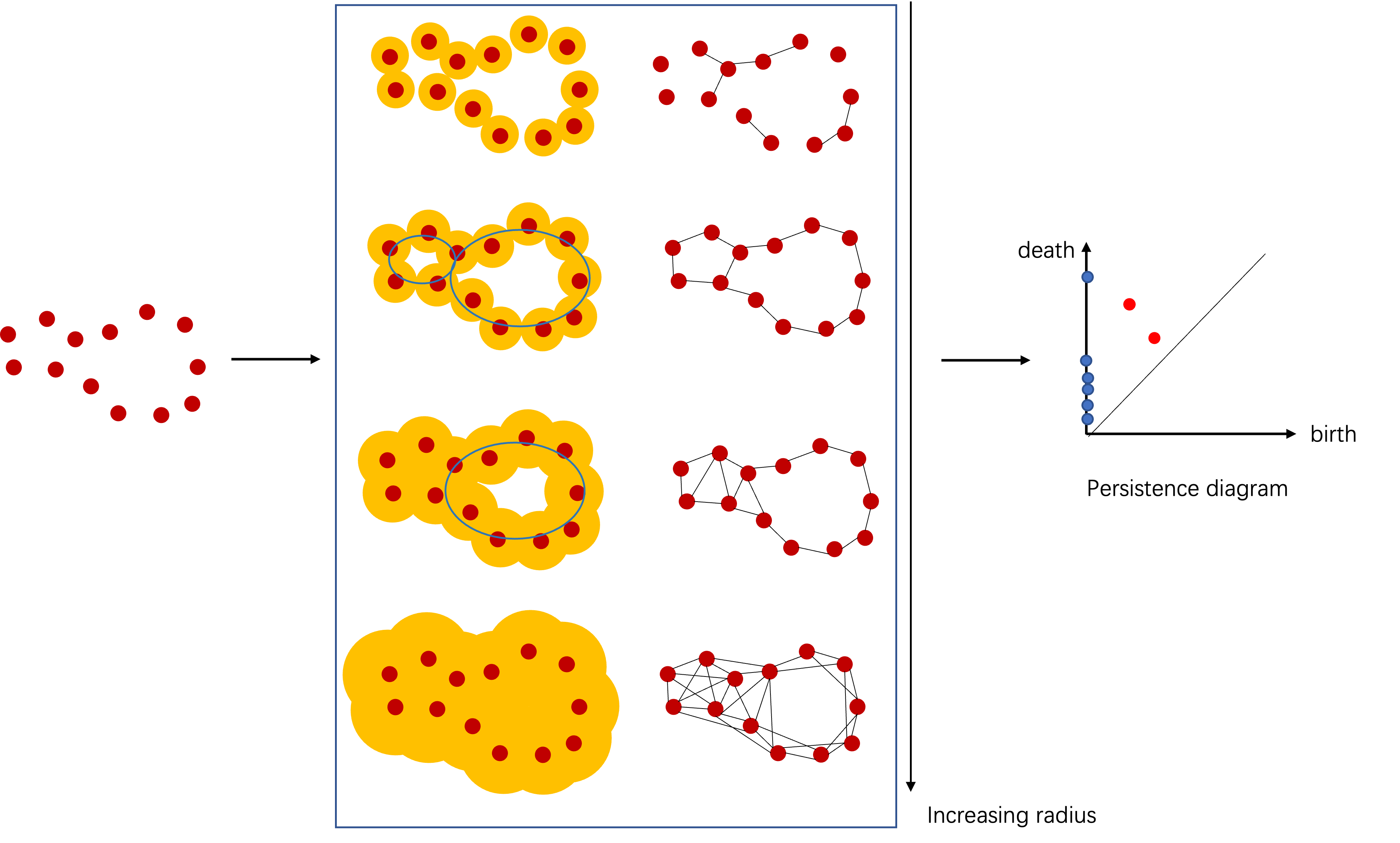}
    \caption{Vietoris-Rips filtration on point clouds. We create a ball centered at each point in the 2D space, and connect two points when their corresponding balls intersect. As radius of balls increase, two voids in the 2D plane appear and then disappear. Their birth and death are recorded as two red persistence points in persistence diagram.}
    \label{fig:vr_complex}
\end{figure}

Given a set of point $V \subseteq \mathbb{R}^n$, the Vietoris-Rips complex at scale $r$ consists of all simplices with diameter less than $r$:
\begin{equation}
    \mathrm{VR}_r (V) = \{\sigma \subset V \mid \forall u, v \in \sigma, ||u - v|| \leq r \}
\end{equation}
In particular, $\mathrm{VR}_0(V) = \{\{u\} || u \in V \}$, and $\mathrm{VR}_{\infty}(V)$ consist of all simplices spanned by vertices in $V$. By increasing $r$ from 0 to $\infty$, we obtain a filtration $\mathrm{VR}_0(V) \subseteq \mathrm{VR}_{r_1}(V) \subseteq \mathrm{VR}_{r_2}(V) \subseteq ... \subseteq \mathrm{VR}_{\infty}(V)$ ($0 \leq r_1 \leq r_2 \leq ...$). We use this Vietoris-Rips filtration on point clouds, see Fig \ref{fig:vr_complex}, to track molecular structures topological features in our experiments.

In order to facilitate the downstream machine learning tasks, a persistence diagram is usually converted a fixed dimensional vector \cite{bubenik2015statistical,Kusano2017Kernel,Carri2017Sliced}. We deploy persistence image \cite{Adams2017Persistence} in our approach, which is a stable and powerful persistence diagrams vectorization approach developed by Adams et al. Set $T: {\reals}^2 \to {\reals}^2$ to be the linear transformation $T(x,y) = (x, y-x)$. Given a persistence diagram $\aD$, let $T(\aD)$ Denote the transformed diagram of a persistence diagram $\aD$ as $T(\aD)$. Let ${\phi}_u:{\reals}^2 \to \reals$ be a differentiable distribution function whose mean locates at $u \in {\reals}^2$: For example, in our implementation later, we will use the Gaussian distribution function (a common choice), where for any $z \in {\reals}^2$, ${\phi}_u(z)=\frac{1}{2 \pi {\tau}^2}e^{-\frac{||z - u||^2}{2 {\tau}^2}}$. 
\begin{definition}[Persistence images]
\label{def:pi}
Let $\alpha : {\reals}^2 \to \reals$ be a non-negative weight function for the persistent plane ${\reals}^2$. Given a persistence diagram $\dgm X$, its \emph{persistence surface} ${\rho}_{\aD}: {\reals}^2 \to \reals$ (w.r.t. $\alpha$) is defined as: for any $z\in \reals^2$, 
\begin{equation}
    \rho_{\aD} (z) = \sum_{u \in T(\aD)} \alpha(u) \phi_u(z). 
\end{equation}

The persistence image is a discretization of the persistence surface as follows. Set a fixed grid within a rectangle in the plane with a collection $\mathcal{P}$ of N pixels. The \emph{persistence image} for a persistence diagram $\aD$ is $\PI_{\aD} = \{ \PI_\aD [p]\}_{p \in \mathcal{P}}$, where $\PI_\aD [p]:=\int \int_{p} \rho_{\dgm(X)} dydx$.
\end{definition}
Note that $\PI_\aD$ can also be viewed as a vector in ${\reals}^N$, and thus persistence images are naturally equipped with the $L_2$-distance in $\reals^N$.  

\section{Graph Neural Networks}
\label{sec:gnn}
Graph Neural Networks (GNN) \cite{gori2005new,scarselli2009graph,bruna2013spectral,niepert2016learning} are the generation of neural networks to the graph structured data. A node in a graph iteratively receives information from its neighborhood and update its representation or features. These node $\&$ edge representation information transferred between vertices are called messages \cite{gilmer2017neural}. A transformation of the messages and updated representations can be learned through training. This message passing scheme can be formulated in a more explicit manner. Given an undirected graph $G = (V, E)$ where $V$ is the node set and $E$ is the edge set, the input to a GNN are node features $h_u^0$ and edge features $e_{uv}$ for every $u \in V$ and $(u,v) \in E$. Then in the $t$-th hidden layer of the GNN, the forward convolution consists of two functions:
\begin{equation}
\label{eq:gnn-framework}
\begin{aligned}
\mathrm{AGGREGATE}~~~~    m^{t+1}_u &= f^t(h^t_u, e_{uv}, \{h^t_v|v \in N(u) \})\\
\mathrm{UPDATE}~~~~    h^{t+1}_u &= g^t(m^{t+1}_u, h^t_u)
\end{aligned}
\end{equation}
where $N(u)$ is the neighborhood of node $u$, $f(\cdot)$ and $g(\cdot)$ are message aggregation and update function respectively. Finally, if the task is on the graph level, there is a readout function $r(\cdot)$ in the final layer mapping node representations to a graph representation
\begin{equation}
    h_G = r(\{h^T_u | u \in V \})
\end{equation}
where $T$ is the total number of hidden layers.

Some popular GNNs include GCN \cite{kipf2016semi}, GraphSAGE \cite{hamilton2017inductive}, GIN \cite{xu2018powerful}, GAT \cite{velivckovic2017graph}, etc. GIN first simply sums the aggregated messages over nodes' neighborhood, and then takes Multiple Layer Perceptrons (MLP) as the the update function. Its forward functions under the formulation of Equation (\ref{eq:gnn-framework}) are:
\begin{equation}
\label{eq:gin}
\begin{aligned}
m^{(t + 1)}_u &= \sum_{v \in N(u)} h_v^{(t)}\\
h^{(t + 1)}_u &= \mathrm{MLP}^{(t)} ((1 + {\epsilon}^{(t)}) h_u^{(t)} + m_u^{(t + 1)})
\end{aligned}
\end{equation}
GAT takes self-attention mechanism to re-weight the messages passed across nodes, and takes a 1-layer MLP to update the aggregated messages. Its message passing functions are:
\begin{equation}
\label{eq:gat}
    \begin{aligned}
    \alpha_{uv}^{(t + 1)} &= \frac{\mathrm{exp}(\mathrm{LeakyReLU}(a^T[W^{(t)} h_u^{(t)}||W^{(t)} h_v^{(t)}]))}{\sum_{v' \in N(u)} \mathrm{exp}(\mathrm{LeakyReLU}(a^T[W^{(t)} h_u^{(t)}||W^{(t)} h_{v'}^{(t)}]))}; \\
    h_u^{(t + 1)} &= \mathrm{ReLU}(\sum_{v \in N(u)} \alpha_{uv}^{(t+1)} W^{(t)} h_v^{(t)} ) . 
    \end{aligned}
\end{equation}
Here, $\alpha_{uv}$ is the attention between nodes $u$ and $v$, $a$ and $W$ are parameters learned from the training process, $(\cdot||\cdot)$ denotes vector concatenation operation, and the two non-linear functions processing features are:
\begin{equation}
    \begin{aligned}
    \mathrm{LeakyReLU}(x) =
    \begin{cases}
    x& \text{x > 0} \\
    \lambda x& \text{x $\leq$ 0}
    \end{cases} ~~\text{and}~~
    \mathrm{ReLU}(x) = 
    \begin{cases}
    x& \text{x > 0} \\
    0& \text{x $\leq$ 0}
    \end{cases}
    \end{aligned}
\end{equation}

Our \HGCN{} takes GIN and GAT layers in our experiments.

\section{More on \HGCN{}}
\label{sec:more_hgcn}
\subsection{$\delta$-Net Clustering}
See Algorithm \ref{alg:clustering} for the detailed $\delta$-Net clustering algorithm used to find hierarchies in CNT bundles for \HGCN{}.

\begin{algorithm}[t]
\caption{$\delta$-Net clustering}
\label{alg:clustering}
\begin{algorithmic}[1]
\REQUIRE A set of points $P=\{p_1, p_2, ..., p_n\}$, radius $\delta > 0$
\ENSURE A $\delta$-net $Q$ of $P$, and a set of clusters $\Pi$ where each point $q\in Q$ corresponding to a cluster in $\Pi$ 
\STATE $P'=\{\}$, $\Pi = \{\}$
\WHILE{$P' \neq $}
\STATE Randomly pick point $p_i \in P \setminus P'$
\STATE Obtain a cluster $C =\{p_j: ||p_i-p_j||_2 \leq \delta|p_j \in P \}$
\STATE $P' = P' \cup C$, $\Pi = \Pi \cup \{C\}$
\ENDWHILE
\STATE Set a node set $V=\{\}$ and an edge set $E=\{\}$
\FOR{$C_i \in \Pi$}
\STATE Create a node $v_i$, add it to $V$
\ENDFOR
\FOR{$(v_i, v_j) \in V \times V$}
\IF{$i \neq j$ and $C_i, C_j$ have overlapping points in P}
\STATE Add edge $\{v_i, v_j\}$ to $E$
\ENDIF
\ENDFOR
\STATE Construct net $Q$ with node set $V$ and edge set $E$
\end{algorithmic}
\end{algorithm}

\subsection{Reweighting factors}
We now describe how to compute the reweighting factor appearing in the message passing process (\ref{eq:re_gcn}), $\tau_{uv}^l$, for two supernodes $u, v\in G^{(k)}$.  
This $\tau_{uv}^l$ is computed by a function taking local geometric information around $u$ and $v$ as input. 

More precisely, the reweighting factors are learned by MLP taking principal and norm vectors mentioned in Section (\ref{subsec:MLpipeline}). In particular, recall given any supernode $a$ in $V_k$ corresponds to a cluster $C_a$ (resp. $C_a$) of nodes from $V_{k-1}$, let $n_a$ denote the approximated normal vector and $\nu_a^i(i=1,2)$ denote two principal vectors at $a$ computed by performing PCA for points in $C_a$. 

Now to compute $\tau_{uv}$, we concatenate information computed from PCA as $\psi_{uv} = n_u || \nu_u^1 || \nu_u^2 || n_v || \nu_v^1 || \nu_v^2$. The rewieghting factor in the $l$-th convolution layer is:
\begin{equation}
    \tau^l_{uv} = S^l(f^l(\psi_{uv})) = \frac{e^{f^l(\psi_{uv})}}{\sum_{v' \in N(u)} e^{f^l(\psi_{uv'})}}
\end{equation}
$f^l$ is a 3-layer MLP and $S^l$ is a softmax function. 

\end{document}